\documentclass[]{amsart}
\usepackage{fullpage}
\usepackage{latexsym}
\usepackage{amsmath}

\usepackage{graphicx}

\usepackage{amsmath}
\usepackage{amssymb}
\usepackage{amsbsy}
\newcommand{\vep}{\varepsilon}

\newcommand{\bT}{{\boldsymbol T}}

\newcommand{\bq}{{\boldsymbol q}}

\newcommand{\fbf}{{\boldsymbol f}}

\newcommand{\bv}{{\boldsymbol v}}

\newcommand{\bx}{{\boldsymbol x}}
\newcommand{\by}{{\boldsymbol y}}

\newcommand{\bX}{{\boldsymbol X}}

\newcommand{\bxi}{{\boldsymbol \xi}}
\theoremstyle{definition}

\theoremstyle{remark}

\numberwithin{equation}{section}

\title{Closure method for spatially averaged dynamics of \\
particle chains}
\author{Alexander Panchenko}
\address{Department of Mathematics, Washington State University, Pullman, WA 99164}
\email{panchenko@math.wsu.edu}
\author{Lyudmyla L. Barannyk}
\address{Department of Mathematics University of Idaho, Moscow, ID 83843}
\email{barannyk@uidaho.edu}
\author{Robert P. Gilbert}
\address{Department of Mathematical Sciences, University of Delaware, Newark, DE 19716}
\email{gilbert@math.edul.edu}

\begin{document}

\bibliographystyle{acm}

\begin{abstract}
We study the closure problem for continuum balance equations that model mesoscale dynamics of large ODE systems.
The underlying microscale model consists of classical Newton equations of particle dynamics. As a mesoscale model we use the balance equations for spatial averages obtained
earlier by a number of authors: Murdoch and Bedeaux, Hardy, Noll and others. The momentum balance equation contains a flux (stress), which is given by an exact function of particle positions and velocities. We propose a method for approximating this function by a sequence of operators applied to average density and momentum. The resulting approximate mesoscopic models are systems in closed form.
The closed from property allows one to work directly with the mesoscale equaitons without the need to calculate underlying particle trajectories, which is useful for modeling and simulation of large particle systems.  The proposed closure method utilizes the theory of ill-posed problems, in particular iterative regularization methods for solving first order linear integral equations. The closed from approximations are obtained in two steps. First, we use Landweber regularization to (approximately) reconstruct the interpolants of relevant microscale quantitites from the average density and momentum. Second, these reconstructions are  substituted into the exact formulas for stress. The developed general theory is then applied to non-linear oscillator chains. We conduct a detailed study of the simplest zero-order approximation, and show numerically that it works well as long as fluctuations of velocity are nearly constant.
\end{abstract}
\maketitle


%%%%%%%
%%%%%%%%%%%%%%%%%%%%%%%%%%%%%%%%%%%%%%%%%%%
%%%%%%%%%%%%%%%%%%%%%%%%%%%%%%%%%%%%%%%%%%%
%%%%%%%%%%%%%%%%%%%%%%%%%%%%%%%%%%%%%%%%%%%%%%%%%%%%%%%%%%%%%
\section{Introduction}
In a series of papers, \cite{mb}, \cite{mb96}, \cite{mb97}, \cite{murdoch07}, Murdoch and Bedeaux studied continuum mechanical balance equations for mesoscopic space time averages of discrete systems. Earlier work of Irving and Kirkwood
\cite{Kirkwood}, Noll \cite{Noll}, and Hardy \cite{Hardy}
on closely related topics
should be also mentioned here.  The fluxes in balance equations (e. g. stress) are given by exact formulas as functions
of particle positions and velocities. This is useful for linking microscale dynamics with mesoscale phenomena. However,
using these formulas requires a complete knowledge of underlying particle dynamics. Since many particle systems of interest have
enormous size, direct simulation of particle trajectories may be intractable.
Consequently, it makes sense to look for closed form approximations of fluxes in
terms of other mesoscale quantities (e.g., average density and velocity), rather than microscopic variables.

In this paper we address the above closure problem for spatially averaged mesoscale dynamics of large size classical
particle chains.
The design of the method was influenced by the following considerations.
\begin{enumerate}
\item The quantities of interest
are space-time continuum
averages, such
as density, linear momentum, stress, energy and others. This choice of averages  is natural because these
quantities are experimentally measurable, and also because of their importance in coupled multiscale simulations involving both
continuum and
discrete models. In addition, by working directly with space-time averages instead of ensemble averages one can bypass a difficult problem of relating
probabilistic and space-time averages.

\item It is desirable to be able to predict behavior of averages on arbitrary time intervals, no matter how short. This
perspective comes from PDE problems, where observation time is often arbitrary
and long time behavior is not of interest. When one tracks an ODE systems on an arbitrary time interval, transients may be
all that is observed. Therefore,
we do not use qualitative theory of ODEs, primarily concerned with describing long time features of dynamics.
This significantly
decreases the range of available tools. However, the closure problem for mesoscopic PDEs
%describing
%evolution of averages
turned out to be a question that can be still answered in a satisfactory way. The methods developed in this fashion
can be helpful in situations where long time features are not of interest: modeling transient and short-lived phenomena, working with metastable
systems, and dealing with problems for which relaxation times can be hard to estimate.
%such as discrete models of multiphase
%continua.

\item We consider particle systems with initial conditions that either known precisely, or ar least such that
the possible initial positions and velocities are strongly restricted by available  a priori information. This is in contrast to
statistical mechanics, where uncertainty of initial conditions is a major problem. In this
regard we note that our approach makes sense for discrete models of solid-fluid continuum systems, where the smallest relevant
length scale is still much larger than a typical intermolecular distance. For other particle systems, our method can be used to run
deterministic simulations repeatedly, in order to accumulate statistical information about
the underlying probability density.

\item
Because of widespread use of
computers in physical and engineering sciences, it is useful to develop theories tailored for computation, rather than
"paper and pencil" modeling.  As far as the closure problem is concerned, traditional phenomenological approach to
formulating constitutive equations can be subsumed by
a more general problem of finding a computational closure method.  In particular, a closure method can be realized as an
iterative procedure where one inputs the values of the primary variables (e.g. density and velocity) computed at the previous
moment of time, and
the algorithm generates the flux (e.g. stress) at the next moment.
Then primary variables  are updated using mesoscopic balance equations, and the process is repeated.
In addition, focusing on computing one can obtain
unconventional but useful continuum mechanical models. By replacing a simple, but possibly crude, Taylor series truncation
with
an algorithm we make it harder
to obtain exact solutions. Since such solutions are rarely available even for simple classical systems, (e.g. Navier-Stokes
equations), this is not a serious drawback. On the positive side, computational closure generally contains an explicit
(explicitly computable) link between micro- and mesoscale properties.

\item An important potential application of closure is development of fast numerical methods for simulating
meso-scopic  dynamics of particle systems. Mesoscale solvers usually employ coarse meshes with mesh size much larger than a typical interparticle distance. Then the averages would be usually given by their coarse mesh values, while interpolants of microscale quantities are discretized on a fine scale mesh. Consequently, a closure method might consist of two generic blocks: (i) reconstruction on mesoscale mesh
thereby a coarse approximations of fine scale quantities are obtained from averages;  and  (ii) interpolation
of the obtained coarse scale discretizations to fine scale.

\end{enumerate}

The closure algorithm developed in the paper is based on iterative regularization methods for
solving first kind integral equations. We observe that primary mesoscale averages are related to the interpolants of
microscale variables via a linear convolution operator. The kernel of this operator
is the "window function" used in \cite{mb} to generate averages. Such integral operators are usually compact.
A compact operator may be invertible, but the inverse operator is not continuous. Therefore, the problem of reconstructing
microscale quantities from given averages is ill-posed. Such problems
are well studied in the literature \cite{Engl2, Gr, Kirsch, Morozov, Tikh1}.
A particular method used in the paper for inverting convolutions is Landweber iteration
\cite{Fridman}, \cite{Land}.  It is known that if
the error
in the data tends to zero, the Landweber method produces successive approximations converging to the exact solution.
For the merely bounded data error, convergence is replaced by a stopping criterion. This criterion provides the optimal number of
iterations needed to approximate the solution with the accuracy proportional to the error in the data. As a consequence, our
method  has desirable feature: one can improve the approximation quality at the price of increasing the algorithm complexity.
This means that predictive capability of the method can be regulated depending on available computing power.

%Previous work. Three groups. One, Murdoch and other papers on space time averaging in classical particle systems.
%Two, work on dimension reduction (Chorin, Majda, E, Van den Einden, Keverekidis, Stuart, Plecac team). Three, relevant
%work on Hamiltonian chains, if any.

The paper is organized as follows. In Section 2 we describe a general multi-dimensional microscopic model. The equations
of motion are classical Newton equations. We limit ourselves to the case of short range interaction forces that may be
either conservative or dissipative. The scaling of particle masses and forces reflects a  continuum mechanical perspective,
that is a family of particle systems of increasing size should represent a hypothetical continuum material.
As $N\to\infty$, the total mass of the system should remain fixed, and the total particle energy should be either fixed,
or at least bounded independent of $N$. Next, we recall the main points of averaging theory of Murdoch-Bedeaux and provide
mesoscopic balance equations and exact formulas for the stress from \cite{murdoch07}.  In Section 3 we develop integral approximations of averages, and describe the use of Landweber iterative regularization for approximate reconstruction. Section 4 contains the formulation of the scaled ODE equations of
the so-called Fermi-Pasta-Ulam (FPU) chains. In Section 5 we derive  closed form
mesoscopic continuum equations of chain dynamics.  The complexity of these continuum models increases with the order $n$ of the iterative deconvolution approximation.
Section 6 is devoted to the detailed study of the simplest closed  model with $n=0$, which we call zero-order closure. Essentially, zero-order closure means that the microscopic quantities are replaced by their averages. Such an approximation can work well only for systems with small fluctuations. To quantify fluctuation size we introduce upscaling temperature and the related notion of quasi-isothermal dynamics. For such dynamics, we show how
to interpolate averages given by mesoscopic mesh values, in order to initialize approximate particle positions and velocities. The interpolation procedure is problem-specific: it conserves microscopic energy and preserves quasi-isothermal nature of the dynamics. Section 7 contains the results of computational tests. Here we apply our
zero-closure algorithm to a Hamiltonian chain with the finite range repulsive potential $U$, decreasing as a power of distance. The results show
good agreement of zero-order approximations with the exact stress produced by direct simulations with 10000-80000 particles, provided the initial conditions have small fluctuations. In our example, the initial conditions are such that  the upscaling temperature is nearly zero during the  observation time. We also demonstrate that  increasing fluctuations of initial velocities leads to a considerable increase in the approximation error, indicating that higher order closure algorithms should be used instead of zero-order closure. Applicability of the zero-order closure is further discussed in Section 8.  Finally, conclusions are provided in Section 9.

\section{Microscale equations and mesoscale spatial averages }
%Microscale problem and mesoscopic mesh
%\subsection{The ODE system}
%\subsection{Multidimensional case}
\subsection{Scaled ODE problems}
The starting point is the microscale ODE problem. In this paper we shall work with classical Newton
equations of point particle dynamics. The same equations may arise as discretization of the
momentum balance equation for continuum systems.
Consider a system containing $N\gg 1$ identical particles, denoted by $P_i$. The mass of each particle
is $\frac{M}{N}$, where $M$ is the total mass of the system. Suppose that during the
observation time $T$, $P_i$ remain inside a bounded domain $\Omega$ in ${\mathbb R}^d$,
where $d$ is the physical space dimension, usually $1, 2$ or $3$.
%\cite{
The positions $\bq_i(t)$ and velocities $\bv_i(t)$ of particles satisfy
 a system of ODEs
\begin{eqnarray}
\dot\bq_i&=&\bv_i,\label{one}\\
 \frac{M}{N}\dot\bv_i&=&\fbf_i+\fbf_i^{(ext)},\label{two}
\end{eqnarray}
subject to the initial conditions
\begin{equation}
\label{inits}
\bq_i(0)=\bx_i, \hspace*{1.0cm} \bv_i(0)=\bv^0_i.
\end{equation}
Here $\fbf_i^{(ext)}$ denotes external forces, such as gravity and confining forces. The interparticle forces $\fbf_i=\sum_j \fbf_{ij}$, where $\fbf_{ij}$ are pair interaction forces which depend on the relative positions and velocities of the respective particles.
%The system (\ref{one}), (\ref{two}) is the {\it microscale model}.

We are interested in investigating asymptotic behavior of the system as $N\to \infty$. Thus it is convenient to introduce
a small parameter
\begin{equation}
\label{small-par}
\vep=N^{-1/d},
\end{equation}
characterzing a typical distance between neighboring particles. As $\vep$ approaches zero, the number of particles goes to infinity, and the distances between neighbors shrink. Consequently,
the
forces in (\ref{two}) should be properly scaled. The guiding principle for scaling is to make
the energy of the system bounded independent of $N$, as $N\to\infty$. In addition, the energy of the
initial conditions should be bounded uniformly in $N$.

As an example of scaling, consider  forces generated by a finite range
potential $U$ and
assume that each particle interacts with no more than a fixed number of neighbors (this is the case, e.g., for particle chains with
nearest neighbor interaction, where a particle always interacts with two neighbors). The fixed number of interacting neighbors
implies that there are about $N$ interacting pairs. Assuming also that the system is sufficiently dense, and variations of
particle concentrations are not large, we can suppose that a typical distance between interacting particles is on the order
$N^{-1/d}L=\vep L$. The resulting scaling
\begin{equation}
\label{force-scaled}
\fbf_{ij}
=-\frac{1}{\vep N}\nabla_{\bx} U\left(\frac{\bq_j-\bq_k}{\vep}\right)
\end{equation}
makes the potential energy of an isolated system bounded independent of $N$. Kinetic energy will be under control
provided the total energy of the initial conditions is bounded independent of $N$.
If exterior forces are present, they should be scaled as well.

\noindent
{\it Remark}. Superficially, the system (\ref{one}), (\ref{two}) looks similar to the parameter-dependent ODE systems studied
in numerous works on ODE time homogenization (see e g. \cite{pavliotis-stuart} and references therein). In the problem under study,
$\vep$ depends on the {\it system dimension} $N$, while in the works on time-homogenization and ODE perturbation theory, the system size is usually fixed as $\vep\to 0$.

%%%%%%%%%%%%%%%%%%%%%%%%%%%%%%%%%%%%%%%%%%%%%%%%%%%%%%%%%%%
\subsection{Length scales}
We introduce the following length scales:\newline
%Our goal is to approximate dynamics of various mesocopic space-time averages (density, velocity, stress etc), which are functionals depending on $\bq_i, \bv_i$ and $\fbf_{ij}$.
\noindent
- macroscopic length scale $L={\rm diam}(\Omega)$;\newline
\noindent
- microscopic length scale
$\vep L$;\newline
\noindent
- mesoscopic length scale $\eta L$,\newline
\noindent
where
$0<\eta< 1$ is a parameter that characterizes spatial mesoscale resolution. This parameter is chosen based on the
desired accuracy, the computational cost requirements, available information about initial conditions and behavior of ODE
trajectories etc.

The computational domain $\Omega$ is subdivided into mesoscopic cells $C_\beta$,
$\beta=1, 2,\ldots,  B$, with the side length on the order of $\eta L$.  The centers $\bx_\beta$ of $C_\beta$ are the nodes of
the meso-mesh. The number of unknowns in the mesoscopic system will be on the order of $B$. For computational efficiency,
one should have $B\ll N$. This does not mean that $\eta$ is close to one. In fact, it makes sense to keep $\eta$ as small as
possible in order to have an additional asymptotic control over the system behavior. Decreasing $\eta$ will in general
make computations more expensive.

%%%%%%%%%%%%%%% FIGURE %%%%%%%%%%%%%%%%%%%%%%%%%%

%\begin{figure}[h]
%\begin{center}
%\includegraphics[width=3in]{out.png}
%\caption{Domain $\Omega$ with particles, mesoscopic mesh, and three length scales.}
%\label{label}
%\end{center}
%\end{figure}

%%%%%%%%%%%%%%%%%%%%%%%%%%%%%%%%%%%%%%%%%%%%%
\subsection{Averages and their evolution}
To define averages we first select a fast decreasing window function
$
\psi
$
satisfying
$
\int \psi(\bx) d\bx=1.
$
There are many possible choices of the window function. In the paper we assume, unless otherwise indicated, that $\psi$ is a compactly supported, differentiable on the interior of its support, and non-negative. Next, define
%For technical reasons $\psi$ must be sufficiently smooth. In practice, we one might use a Gaussian, or a smooth compactly supported cut-off function. Next, set
$$
\psi_\eta(\bx)=\eta^{-d} \psi\left(\frac{\bx}{\eta}\right).
$$

%\subsection{Mesoscopic averages}
Once the window function is chosen, we can evaluate the averages of various continuum mechanical variables, following \cite{mb}, \cite{murdoch07}.
%The total number and mass of particles inside $C_j$ is well approximated by, respectively,
%$$
%\sum_i \psi_\eta(\bx_\beta-\bq_i(t)),
%$$
%and
%$$
%\sum_i m_i\psi_\eta(\bx_\beta-\bq_i(t)),
%$$
%where the summation is over all particles.
%The function $\psi^\eta$ acts as filter, counting only particle inside the support of $\psi^\eta_i$. Most of these particles a counted with the weight 1, except particles close to the boundary of $C_j$. The volume of $C_j$ is approximated by $V_j=\int \psi^\eta =c \eta^d$, where $d$ denotes the dimension of the underlying physical space (1, 2, or 3), and $c$ is independent of $\eta$.
%Dividing the total mass of particles
%inside $C_j$  by $V_j$
 The mesoscopic average density and momentum are given by
\begin{equation}
\label{density}
\overline{\rho}^\eta(t, \bx)=\frac{M}{N}\sum_{i=1}^N \psi_\eta(\bx-\bq_i(t)),
\end{equation}
%Similarly, we can approximate the spatially averaged momentum,
%%%%%%%%%%%%%%%%%%%%%%%%%%%%%%%%%%%%%%%
\begin{equation}
\label{mom}
\overline{\rho}^\eta \overline{\bv}^\eta(t, \bx)=\frac{M}{N}\sum \bv_i(t)\psi_\eta(\bx-\bq_i(t)).
\end{equation}
The meaning of the above definitions becomes clear if one considers $\psi=(c_d)^{-1} \chi(x)$, where $\chi$ is a characteristic function of the unit ball in ${\mathbb R}^d$, and $c_d$ is the volume of the unit ball. Then
$$
\overline{\rho}^\eta=\frac{1}{c_d \eta^d}\frac{M}{N} \sum \chi\left(\frac{\bx-\bq_i(t)}{\eta}\right).
$$
The sum in the right hand side gives the number of particles located within distance $\eta$ of $\bx$ at time $t$.
Multiplying by $M/N$ we get the total mass of these particles, and dividing by $c_d \eta^d$ (the volume of $\eta$-ball) gives
the usual particle density.

Differentiating (\ref{density}), (\ref{mom}) in $t$, and using the ODEs (\ref{one}), (\ref{two}) one can obtain \cite{mb}
exact mesoscopic balance equations for all primary variables. For example, for an isolated system with ($\fbf_i^{(ext)}=0$), mass conservation and momentum balance equations take the form:
\begin{equation}
\label{mass-balance}
\partial_t \overline{\rho}^\eta+{\rm div}(\rho^\eta\overline{\bv}^\eta)=0,
\end{equation}
%%%%%%%%%%%%%%%%%%%%%%%%%%%%%%%%%%%%%%%%%%%
\begin{equation}
\label{m-balance}
\partial_t(\overline{\rho}^\eta\overline{\bv}^\eta)+{\rm div}\left(\overline{\rho}^\eta\overline{\bv}^\eta\otimes
\overline{\bv}^\eta\right) - {\rm div}\bT^\eta=0.
\end{equation}
The stress
$\bT^\eta=\bT^\eta_{(c)}+\bT^\eta_{(int)}$ \cite{murdoch07}, where
\begin{equation}
\label{m-stress-c}
\bT^\eta_{(c)}(t, \bx)=-\sum m_i(\bv_i-\overline{\bv}^\eta(t, \bx, ))\otimes (\bv_i-\overline{\bv}^\eta(\bx, t))\psi(\bx-\bq_i)
\end{equation}
is the {\it convective stress}, and
\begin{equation}
\label{m-stress-int}
\bT^\eta(t, \bx)_{(int)}=
\sum_{(i, j)}\fbf_{ij}\otimes (\bq_j-\bq_i)\int_0^1 \psi\left(s(\bx-\bq_j)+(1-s)(\bx-\bq_i)\right)ds
\end{equation}
is the {\it interaction stress}. The summation in (\ref{m-stress-int}) is over all pairs of particles $(i, j)$ that interact with each other.

Discretizing balance equations on the mesoscopic mesh yields a discrete system of equations, called the {\it meso-system}, written for mesh values
of $\overline{\rho}^\eta_\beta, (\overline{\rho}^\eta\overline{\bv}^\eta)_\beta$ and $\bT^\eta_\beta$.
The dimension of the meso-system is much smaller than the dimension of the original ODE problem. However, at this stage we still have no computational savings, since the meso-system is {\it not closed}. This means that mesoscopic fluxes such as
(\ref{m-stress-c}), (\ref{m-stress-int}) are expressed as functions of the microscopic positions and velocities. To find these
positions and velocities, one has to solve the original microscale system (\ref{one}), (\ref{two}).
To achieve computational savings we need to replace exact fluxes with approximations that involve only mesoscale quantities.
We refer to the procedure of generating such approximations as a {\bf closure method}.
This closure-based approach has much in common with continuum mechanics.
The important difference is that the focus is on computing, rather than continuum mechanical style
modeling of constitutive equations.
%%%%%%%%%%%%%%%%%%%%%%%%%%%%%%%%
%%%%%%%%%%%%%%%%%%%%%%%%%%%%%%%%%
\section{Closure via regularized deconvolutions}
\subsection{Outline}
Our approach is based on a simple idea: the integral approximations of primary averages (such as density and velocity) are related to the corresponding microscopic quantities via convolution with the kernel $\psi_\eta$.  Therefore, given primary variables we can (approximately) recover the microscopic positions and velocities by numerically inverting convolution operators. The results are inserted into equations for secondary averages (or fluxes), such as stress in the momentum balance. This yields closed form balance equations that can be simulated efficiently on the mesoscopic mesh.
%%%%%%%%%%%%%%%%%%%%%%%%%%%%%%%%%%%%%%%%
\subsection{Integral approximation of discrete averages}
\label{sect:int-appr}
To exploit the special structure of primary averages, it is convenient to approximate sums such as
\begin{equation}
\label{gen-av}
\overline{g}^\eta=\frac{1}{N}
\sum_{j=1}^Ng(\bv_j, \bq_j) \psi_\eta(\bx-\bq_j)
\end{equation}
by integrals.
Since particle positions $\bq_j$ are not periodically spaced, (\ref{gen-av}) is not in general a Riemann sum for
$g \psi_\eta(\bx-\cdot)$. To interpret the sum correctly, we introduce interpolants
$\tilde{\bq}(t, \bX), \tilde{\bv}(t, \tilde{\bq})$ of positions and velocities, associated with the microscopic ODE system
(\ref{one}), (\ref{two}). At $t=0$ these interpolants satisfy
$$
\tilde{\bq}(0, \bX_j)=\bq_j^0,~~~~~~~~~~\tilde{\bv}(0, \tilde{\bq}(0, \bX_j))={\bv}^0_j,
$$
where $\bX_j$, $j=1, 2, \ldots, N$ are points of $\vep$-periodic rectangular lattice in $\Omega$.
At other times,
$$
\tilde{\bq}(t, \bX_j)=\bq_j(t), ~~~~~~~\tilde{\bv}(t, \tilde{\bq}(t, \bX_j))=\bv_j(t).
$$

%For any choice of initial conditions $\bX, \bV_0$, the functions $\tilde{\bQ}(t, \cdot), \tilde{\bV}(t, \cdot)$ provide a solution of (\ref{three}) satisfying the initial condition
%$$
%\bQ(0)=\bX, ~~~~\bV(0)=\bV_0.
%$$
%We also define a positions-to-velocities map
%\begin{equation}
%\label{w}
%\bw(\tilde{\bq}(t))=\tilde{\bv}(t, \bX).
%\end{equation}
%In practice, $\bQ$ and $\bW$ can be approximated by interpolating microscopic positions and velocities at each time $t$.
%We will identify these maps with the corresponding interpolants, to simplify notations.
%Next, suppose that at the initial time, particles are placed at the points $\bX_j, j=1, 2, \ldots, N$ be  on $\vep$-periodic mesh in $\Omega$.
Then we can rewrite (\ref{gen-av}) as
\begin{equation}
\label{gen-av2}
\overline{g}^\eta=\frac{1}{|\Omega|}
\sum_{j=1}^N \frac{|\Omega|}{N} g
\left(
\tilde{\bv}
\left(t, \tilde{\bq}(t, \bX_j)
\right),
\tilde{\bq}(t, \bX_j
\right)
\psi_\eta(\bx-\tilde{\bq}(t, \bX_j)),
\end{equation}
where $|\Omega|$ denotes the volume (Lebesgue measure) of $\Omega$.
Eq. (\ref{gen-av2})  is a Riemann sum generated by partitioning $\Omega$ into $N$ cells of volume $|\Omega|/N$
centered at $\bX_j$. This yields
\begin{equation}
\label{int1}
\overline{g}^\eta =\frac{1}{|\Omega|} \int_\Omega g
\left(
\tilde{\bv}(t, \tilde{\bq}(t, \bX)), \tilde{\bq}(t, \bX)
\right)
\psi_\eta (\bx-\tilde{\bq}(t, \bX)) d\bX,
\end{equation}
up to discretization error.
Now suppose that the map $\tilde{\bq}(\cdot, \bX)$ is invertible for each $t$, that is $\bX=\tilde{\bq}^{-1}(t, \tilde{\bq})$. Changing the variables in the integral $\by=\tilde{\bq}(t, \bX)$ we obtain a generic integral approximation
\begin{equation}
\label{int2}
\overline{g}^\eta= \frac{1}{|\Omega|} \int_\Omega g\left(\tilde{\bv}(t, \by), \by
\right) \psi_\eta (\bx-\by) J(t, \by)~d\by,
\end{equation}
where
\begin{equation}
\label{J}
J=|\det \nabla \tilde{\bq}^{-1}|,
\end{equation}
up to discretization error.

\subsection{Regularized deconvolutions}
Define an operator $R_\eta$ by
$$
R_\eta[f](\bx)=\int \psi_\eta(\bx-\by) f(\by) d\by.
$$
To simplify exposition, suppose that $R_\eta$ is injective. For example, a Gaussian $\psi_\eta$ produces an injective operator,
which is not difficult to check using Fourier transform and uniqueness of analytic continuation.
If $R_\eta$ is injective, then there exists the single-valued
inverse operator $R^{-1}_\eta$, that we call the {\it deconvolution operator}. Unfortunately, this operator is unbounded, since
$R_\eta$ is compact in $L^2(\Omega)$. This is the underlying reason for
the popular belief that averaging destroys the high-frequency information contained in the microscopic quantities.
In fact, this information is still there (the inverse operator exists), but it is difficult to recover in a stable manner, because of unboundedness. This does not make the situation hopeless, as has been recognized for some time. Reconstructing $f$ from the knowledge of $R_\eta[f]$) is a classical example of an unstable ill-posed problem
(small perturbations of the right hand side may
produce large perturbations of the solution). The exact nature of ill-posedness and methods of regularizing the problem are well investigated both analytically and numerically (see, e. g.  \cite{Gr, Kirsch, Morozov, Tikh1, Engl1, Engl2}).
Accordingly, we interpret notation
$R_\eta^{-1}$ as a suitable regularized approximation of the exact
operator. Many regularizing techniques are currently available: Tikhonov regularization, iterative methods, reproducing kernel methods, the maximum entropy method, the dynamical system approach and others. It is very fortunate that this vast array of knowledge can be used for the ODE model reduction.
On the conceptual level, our approach makes it clear that instability associated with ill-posedness is a fundamental difficulty in the process of closing the continuum mechanical equations.

A family of Landweber iterative deconvolution methods \cite{Fridman}, \cite{Land}  seems to be particularly convenient in the present context. In the simplest version, approximations $g_n$ to the solution of the
operator equation
\begin{equation}
\label{op-eq}
R_\eta[g]=\overline{g}^\eta
\end{equation}
are generated by the formula
\begin{equation}
\label{seq1}
g_n=\sum_{k=0}^n (I-R_\eta)^n \overline{g}^\eta, \;\;\;\; g_0=\overline{g}^\eta.
\end{equation}
The number $n$ of iterations plays the role of regularization parameter.
In (\ref{seq1}), $I$ denotes the identity operator.

%INSERT REFS ABOUT CONVERGENCE, ESTIMATES AND STOPPING CRITERIA.

The first three low-order approximations are
 \begin{eqnarray}
&  g_0= \overline{g}^\eta & n=0,\label{pract-appr0}\\
& g_1 = \overline{g}^\eta+(I-R_\eta)[\overline{g}^\eta]& n=1, \label{pract-appr1}\\
& g_2=\overline{g}^\eta+(I-R_\eta)[\overline{g}^\eta]+(I-R_\eta)^2[\overline{g}^\eta]& n=2.\label{pract-appr2}
\end{eqnarray}

\section{Microscale particle chain equations}
In this section, the general method outlined above is detailed in the case of one-dimensional Hamiltonian chain of oscillators that consists of $N$ identical particles with nearest neighbor interaction. The domain $\Omega$ is an interval $(0, L)$.
Particle positions, denoted by $q_j=q_j(t)$, $j=1,\ldots, N$, satisfy
\[
0<q_1<q_2<\ldots<q_N<L
\]
at all times, i.e. the particles cannot occupy the same position or jump over each other. Next, define a small
parameter
$$
\vep=\frac{1}{N},
$$
and microscale step size
\begin{equation}
\label{h}
h=\frac{L}{N}.
\end{equation}

The interparticle forces
\begin{equation}
\label{micro-force}
f_{jk}=
\frac{q_j-q_k}{|q_j-q_k|}
U^{\prime}\left(\frac{|q_j-q_k|}{\vep}\right)
\end{equation}
are defined by a finite range potential $U$. We suppose that $U^\prime(\xi)\geq 0$ for all $\xi$ within the range.
Note that $k$ in (\ref{micro-force}) can take only two values: $j-1$ or $j+1$. Also, observe
$f_{jk}=-f_{kj}$, as it should be by the third law of Newton, and also that the sign of $f_{jk}$ is the same as sign of $q_j-q_k$. This means that the force exerted on $P_j$ by say, $P_{j+1}$ is repulsive.
The total interaction force acting on the particle $P_j$ is
$$f_j=f_{j, j-1}+f_{j,j+1},$$
for $j=2, 3, \ldots, N-1$.

Each particle has mass $m=M/N=M\vep$, where $M$ is the total mass of the system.  Particles have velocities denoted by $v_j$, $j=1,\ldots,N$.
Writing the second Newton's law as a system of first order equations yields the scaled microscale ODE system
\begin{equation}
\label{ode-1}
\begin{array}{l}
\dot{q_j}=v_j, ~~~ \vep M\dot{v_j}=f_j, \;\;\; j=1,\dots,N
\end{array}
\end{equation}
subject to the initial conditions
\begin{equation}
\label{ode-2}
q_j(0)=q_j^0, \;\;\; v_j(0)=v_j^0.
\end{equation}
%%%%%%%%%%%%%%%%%%%%%%%%%%%%%%%%%%%%%%%%%%%%%%%
%%%%%%%%%%%%%%%%%%%%%%%%%%%%%%%%%%%%%%%%%%%%%%%
\section{Integral approximation of stresses for particle chains. Mesoscopic continuum equations}
In the one-dimensional case stress is a scalar quantity, and (\ref{m-stress-c}), (\ref{m-stress-int}) reduce to, respectively,
\begin{equation}
\label{c1}
T^\eta_{(c)}(t, x)=-\sum_{j=1}^N \frac{M}{N}
(
v_j-\overline{v}^\eta(t, x)
)^2
\psi_\eta (x-q_j),
\end{equation}
and
\begin{equation}
\label{c2}
T^\eta_{(int)}(t, x)=
\sum_{j=1}^{N-1} f_{j, j+1} (q_{j+1}-q_j)\int_0^1 \psi_\eta(x-s q_{j+1}-(1-s)q_j)ds.
\end{equation}
The sum in (\ref{c2}) is simplified compared to the general expression, since we have exactly $N-1$ interacting pairs of particles.

To obtain integral approximations of stresses, we define interpolants $\tilde q, \tilde v$, as in Sect. \ref{sect:int-appr}.
Assuming as before that $\tilde q$ is invertible and repeating the calculations we get
\begin{equation}
\label{c3}
T^\eta_{(c)}(t, x)=
-\frac{M}{L} \int_0^L \left(
\tilde v(t, y)-\overline{v}^\eta(t, x)
\right)^2
\psi_\eta(x-y) J(t, y) dy.
\end{equation}

\noindent
{\it Remark}. Many equalities in the paper, including (\ref{c3}) hold up to a discretization error. To simplify presentation, we do not mention this in the sequel when discrete sums are approximated by integrals.

The interaction stress can be rewritten as
\begin{equation}
\label{c4}
T^\eta_{(int)}(t, x)=-
\frac{N-1}{N}
\sum_{j=1}^{N-1}\frac{L}{N-1} U^\prime
\left(
\frac{q_{j+1}-q_j}{h}L
\right)
 \frac{q_{j+1}-q_j}{h}\int_0^1 \psi_\eta(x-s q_{j+1}-(1-s)q_j)ds.
\end{equation}
Next we approximate $(q_{j+1}-q_j)/h$ by $\tilde q^\prime(t, X_j)$. This approximation is in fact exact, provided the interpolant is chosen to be piecewise linear. Note also that
$$
\tilde q^\prime(t, X)=\frac{1}{(\tilde q^{-1})^\prime(t, \tilde q(t, X))}=\frac{1}{J(t, \tilde q(t, X))}.
$$
Inserting this into (\ref{c4}), replacing Riemann sum with an integral and changing variable of integration
as in Sect. \ref{sect:int-appr}, we obtain the integral approximation of the interaction stress:
\begin{equation}
\label{c5}
T^\eta_{(int)}(t, x)=-
\int_0^L U^\prime
\left(
\frac{L}{J(t, y)}
\right)
\int_0^1 \psi_\eta
\left(
x-y-\frac{sh}{J(t, y)}
\right)
ds~
dy.
\end{equation}
Equations (\ref{c3}), (\ref{c5}) contain two microscale quantities: $J$ and $\tilde v$. Approximating sums in
the definitions of the primary averages (\ref{density}), (\ref{mom}) by integrals we see that $\overline{\rho}^\eta$ and $\overline{v}^\eta$ are obtained by applying the convolution operator $R_\eta$ to, respectively
$J$ and $J \tilde v$:
\begin{equation}
\label{c6}
\overline{\rho}^
\eta=\frac{M}{L} R_\eta [J],\hspace*{1.5cm}\overline{\rho}^
\eta\overline{v}^\eta=\frac{M}{L} R_\eta [J \tilde v].
\end{equation}
The discretization error in (\ref{c6}) can be made small by imposing suitable requirements on the microscopic interpolants. Fortunately, the theory of ill-posed problems allows for errors in the right hand side of integral equations. The size of the error determines the choice of regularization parameter. In the present case, the error determines the number of iterations needed for the optimal reconstruction, according to so-called stopping criteria. These criteria are available in the literature on ill-posed porblems (see e.g. \cite{Kirsch}). Detailed investigation of these
questions is left to future work.

Denote by $R_{\eta, n}^{-1}$ the iterative Landweber regularizing operators
$$
R_{\eta, n}^{-1}=\sum_{k=0}^n (I-R_\eta)^k.
$$
Applying $R_{\eta, n}^{-1}$ in (\ref{c6}) yields a sequence of approximations
\begin{equation}
\label{c7}
J_n= \frac{L}{M} R_{\eta, n}^{-1} [\overline{\rho}^
\eta], \hspace*{1.5cm}~~~~~~~~~ ~~~~~~\tilde v_n= \frac{R_{\eta, n}^{-1} [\overline{\rho}^
\eta \overline{v}^\eta]}{R_{\eta, n}^{-1} [\overline{\rho}^
\eta]},
\end{equation}
and a corresponding sequence of closed form mesoscopic continuum equations (written here for an isolated system with zero exterior forces)
\begin{eqnarray}
\partial_t \overline{\rho}^\eta+\partial_x (\overline{\rho}^\eta\overline{v}^\eta) &=& 0,\label{c8}\\
\partial_t( \overline{\rho}^\eta\overline{v}^\eta)+\partial_x
\left(
\overline{\rho}^\eta(\overline{v}^\eta)^2
\right)
-\partial_x (T^\eta_{(c), n}+T^\eta_{(int), n})& =& 0,\label{c9}
\end{eqnarray}
where $T^\eta_{(c), n}, T^\eta_{(int), n}$ are given by
\begin{equation}
\label{c10}
T^\eta_{(c), n}=
-\frac{M}{L} \int_0^L \left(
\tilde v_n(t, y)-\overline{v}^\eta(t, x)
\right)^2
\psi_\eta(x-y) J_n(t, y) dy,
\end{equation}
\begin{equation}
\label{c11}
T^\eta_{(int), n}=-
\int_0^L U^\prime
\left(
\frac{L}{J_n(t, y)}
\right)
\int_0^1 \psi_\eta
\left(
x-y-\frac{sh}{J_n(t, y)}
\right)
ds~
dy,
\end{equation}
with $J_n, \tilde v_n$ given by (\ref{c7}).

\section{Zero-order closure for particle chains}
\label{sect:zero-closure}
Let us consider zero-order approximations in detail.
The mesoscopic mesh consists of points
$$
x_\beta=\left(\beta-\frac 12\right)\eta L,~~~ \beta=1, 2,\ldots, B,
$$
where $B=1/\eta$, presumed to be an integer satisfying $B\ll N$. Meso-cells are intervals $I_\beta$ of length
$$L_\eta=L/B=\eta L,$$ centered at $x_\beta$.

Suppose
that the only primary variables of interest are density
$\overline{\rho}^\eta$ and linear momentum $\overline{\rho}^\eta\overline{v}^\eta$.
These variables will be computed by the mesoscale solver. For simplicity, suppose that the meso-solver is explicit in time.
Then the average density and average velocity will be available at the previous moment of time. Our task is to design an
update step for computing density and velocity at the next time moment.  To construct a closed form update step, we need to
approximate stress $T^\eta$ in (\ref{m-balance}) in terms of $\overline{\rho}^\eta, \overline{\rho}^\eta\overline{v}^\eta$.
From the knowledge of $\overline{\rho}^\eta, \overline{\rho}^\eta\overline{v}^\eta$ we can approximately recover
$J$ and $J\tilde{v}$.  The zero-order approximation (\ref{pract-appr0}) corresponds to
\begin{eqnarray}
\label{0-1}
J(t, x) & \approx &\frac{L}{M} \overline{\rho}^\eta (t, x),\\
J(t, x) \tilde{v}(t, x)& \approx  & \frac{L}{M} \overline{\rho}^\eta \overline{v}^\eta (t, x)
\label{0-2}.
\end{eqnarray}
In other words, the microscale quantities are approximated by their averages. The corresponding closed form approximations
for stress are obtained by
inserting (\ref{0-1}), (\ref{0-2}) into (\ref{c10}), (\ref{c11}):
\begin{equation}
\label{0-8}
T^\eta_{(c), 0}(t, x)=
-\int_0^L \left(
\overline{v}^\eta(t, y)-\overline{v}^\eta(t, x)
\right)^2
\psi_\eta(x-y) \overline{\rho}^\eta(t, y) dy,
\end{equation}
\begin{equation}
\label{0-81}
T^\eta_{(int), 0}=-
\int_0^L U^\prime
\left(
\frac{M}{\overline{\rho}^\eta(t, y)}
\right)
\int_0^1 \psi_\eta
\left(
x-y-\frac{sh M}{L \overline{\rho}^\eta(t, y)}
\right)
ds~
dy.
\end{equation}
For computation, a numerical quadrature should be used. In this regard, note that all average quantities are computed on the
mesoscale mesh, while the formulas (\ref{m-stress-c}), (\ref{m-stress-int}) are fine scale discretizations. Therefore, one
might wonder if a straightforward mesoscale quadrature of (\ref{0-8}), (\ref{0-81}) is too crude. A better approach is
to interpolate $\overline{\rho}^\eta, \overline{v}^\eta$ by
prescribing approximate particle positions $\hat q_j$ and velocities $\hat v_j$, $j=1, 2, \ldots, N$, compatible with the
given $\overline{\rho}^\eta, \overline{v}^\eta$. Once this is done, (\ref{0-8}), (\ref{0-81}) can be discretized on a fine
scale mesh with mesh nodes $\hat q_j$.

Interpolants cannot be unique. For zero-order closure, we
are choosing positions and velocities that produce the given average density and average velocity. Clearly, there are
many different position-velocity configurations with the same averages. The choice made in the paper is motivated by
the practical requirement of achieving low operation count, as well as by certain expectations about the nature of dynamics.
From the continuum mechanical point of view, if a system
can be adequately modeled by balance equations of mass and momentum, then it must have have a trivial energy balance.
Most often this means that the
deformation is nearly isothermal. To mimic such isothermal dynamics we suppose that at each time step, there exists a positive
number $\kappa^2$ (it can be called "upscaling temperature") such that
\begin{equation}
\label{0-4}
\sum_{j\in J_\beta} (v^\beta_j-\overline{v}^\eta(t, x_\beta))^2\psi_\eta(x_\beta-\hat q_j)=\kappa^2.
\end{equation}
Here the summation is over all particles located in a meso-cell $I_\beta$. The temperature $\kappa^2$ is the same for all
$\beta=1, 2, \ldots, B$. We emphasize that {\it the actual value} of $\kappa^2$ is not as important as the fact that its value
is the same for all meso-cells. This is because (\ref{0-4}) would yield
constant mesoscale mesh node values of $T^\eta_{(c), 0}$ As a result, the finite difference approximation of
$\partial_x T^\eta_{(c), 0}$ on the mesoscale mesh is identically zero. We interpret this by saying that
convective stress does not contribute to the mesoscopic dynamics in the isothermal case. Another observation is that
$\kappa$ need not be the same at different moments of time, so our assumption is somewhat more flexible than the standard
isothermal deformation approximation. Also, we note that validity (\ref{0-4}) depends on the choice of $\eta$. For bigger
$\eta$, it is more likely that (\ref{0-4}) holds for the same underlying microscopic dynamics.
Details on this are provided below in Section \ref{sect:velocities}.
Additionally, other features of the microscopic dynamics should be taken into account. Most importantly, interpolated velocities
$\hat v_j=\overline{v}^\eta (\hat q_j)$ must be such that the collection $\hat q_j, \hat v_j, j=1, 2, \ldots, N$
conserves microscopic energy ${\mathcal E}$:
\begin{equation}
\label{en-cons}
{\mathcal E}=\frac 12 \frac{M}{N} \sum_{j=1}^N (\hat{v}_j)^2+{\mathcal U}(\widehat{Q})
\end{equation}
where ${\mathcal U}(\widehat{Q})$ is the microscale potential energy corresponding to the positions $\hat q_j$.

\subsection{Prescribing particle positions}
The objective of this section is to assign approximate particle positions $\hat q_j$.
%In general,
%$\hat q_j$ are different from the actual particle positions $q_j$.
%The choice of $\hat q_j$ must be compatible with approximations of $\tilde v, J$ obtained from zero-order closure.
%In practice, these approximations will be computed at the mesh nodes $x_\beta, \beta=1, 2, \ldots, B$.
%Therefore, interpolation is needed before $\hat q_j$ can be assigned.
We start by interpolating $J$. The simplest interpolant is piecewise-constant:
$
J(t, \bx) \approx \sum_{\beta=1}^B J(t, x_\beta)\chi_\beta(\bx)=
\sum_{\beta=1}^B\frac{L}{M} \overline{\rho}^\eta (t, x_\beta) \chi_\beta(\bx)
$
where $\chi_\beta$ is the characteristic function of the meso-cell $I_\beta$.  A simple choice of the position map
compatible with this interpolant is a piecewise linear map having the prescribed constant value of $J$ in each meso-cell.
In practical terms, this means that in each meso-cell, particles are spaced at equal intervals from each other. The local
interparticle spacing
\begin{equation}
\label{dbt}
\Delta_\beta=\frac{M}{\overline{\rho}^\eta(t, x_\beta) N}
\end{equation}
is determined by
the mesh value of the average density. To explain (\ref{dbt}), note that the total mass of particles contained in the
meso-cell $I_\beta$ can be approximated by $\overline{\rho}^\eta(t, x_\beta) L_\eta$.
Dividing by the mass $M/N$ of one particle, we obtain an approximate number of particles inside $I_\beta$:
$$
n_\beta=\overline{\rho}^\eta(t, x_\beta) L_\eta\frac{N}{M},
$$
and thus $\Delta_\beta=L_\eta/n_\beta$.
%In the sequel, we denote by $\widehat{Q}$ the
%collection of particle positions assigned according to the above prescription.
We emphasize that $\hat q_j$ are chosen based only on the known mesh values of the density $\overline{\rho}^\eta$,
and that $\hat q_j$ will be different from the actual particle positions $q_j$.

Now we approximate the integral in (\ref{0-81}) by its Riemann sum generated by the partition $\{\hat q_j, j=1, 2, \ldots, N\}$:
\begin{eqnarray}
\label{0-71}
T^\eta_{(int), 0} & \approx & -
\sum_{j=1}^{N-1}
 U^\prime
\left(
N (\hat q_{j+1}-\hat q_j)
\right)
(\hat q_{j+1}-\hat q_j)
\int_0^1 \psi_\eta
\left(
x-s \hat q_{j+1}-(1-s)\hat q_j
\right)
ds.
\end{eqnarray}
%%%%%%%%%%%%%%%%%%%%%%%%%%%%%%%%%%%%%%%%%%%%%%%%%%%%%%
%%%%%%%%%%%%%%%%%%%%%%%%%%%%%%%%%%%%%%%%%%%%%%%%%%%%%%
\subsection{Prescribing particle velocities}
\label{sect:velocities}
In order to approximate the convective stress in the fine scale discretization of (\ref{0-8}), we need to choose approximations $\hat v_j$ of the true particle
velocities $v_j$. The choice of $\hat v_j$ must satisfy (\ref{en-cons}) and
be compatible with the available average velocity at the mesoscale mesh nodes.

For each $\hat q_j\in I_\beta$, we set
$$
\hat v_j=\overline{v}^\eta_\beta+\delta v^\beta_j,
$$
where $\overline{v}^\eta_\beta$ is the local average velocity, and $\delta v^\beta_j$ is a perturbation to be defined.

Next, we show that the energy-conserving collection of $\delta v_j$ velocity always exists,
provided its upscaling temperature is suitably prescribed. This prescription will be based only on the available
mesoscale information.
For definitiveness, in the rest of this section we suppose that $\psi_\eta$ satisfies the
following condition:
\begin{eqnarray}
\label{c11-1}
\psi_\eta(x_\beta-\hat q_j)>0, ~{\rm if}~\hat q_j\in I_\beta.&&
%\psi_\eta~{\rm is}~{\rm even},&& \label{c2} \\
%\psi_\eta~{\rm has}~{\rm a}~{\rm maximum}~{\rm at}~0.&& \label{c3}
\end{eqnarray}
To make the algebra simpler, we make another assumption: for each $\beta=1, 2, \ldots, B$,
\begin{equation}
\label{simpl}
\sum_{j=1}^N
f(\hat v_j)\psi_\eta(x_\beta-\hat q_j)
\approx
\sum_{j\in J_\beta}
f(\hat v_j) \psi_\eta(x_\beta-\hat q_j),
\end{equation}
where $f$ is either $\hat v_j$ or $(\hat v_j)^2$. The second summation is over all $j$ such that $\hat q_j\in I_\beta$.
Assumption (\ref{simpl}) holds when $\psi_\eta(x_\beta-y)$ is small outside of $I_\beta$.

Averaging of  $\hat v_j$ should produce the known
average velocity
$\overline{v}^\eta_\beta$. This yields an equation for $\delta v_j$:
\begin{equation}
\label{0-9}
\frac{M}{N} \sum_{j\in J_\beta}\hat v_j\psi_\eta(x_\beta-\hat q_j)=\overline{\rho}^\eta_\beta\overline{v}^\eta_\beta.
\end{equation}
Since
\begin{eqnarray*}
\frac{M}{N} \sum_{j\in J_\beta}\hat v_j \psi_\eta(x_\beta-\hat q_j)& = &
\overline{v}^\eta_\beta \frac{M}{N} \sum_{j\in J_\beta}\psi_\eta(x_\beta-\hat q_j)+
\frac{M}{N} \sum_{j\in J_\beta}\delta v^\beta_j \psi_\eta(x_\beta-\hat q_j)
 \\
& = &
\overline{\rho}^\eta_\beta \overline{v}^\eta_\beta+
\frac{M}{N} \sum_{j\in J_\beta}\delta v^\beta_j \psi_\eta(x_\beta-\hat q_j),
\end{eqnarray*}
(\ref{0-9}) holds provided
\begin{equation}
\label{0-10}
\frac{M}{N} \sum_{j\in J_\beta}\delta v^\beta_j \psi_\eta(x_\beta-\hat q_j)=0.
\end{equation}
Now we look for perturbations in the form
\begin{equation}
\label{0-31}
\delta v_j^\beta=\frac{a_j^\beta}{ \psi_\eta(x_\beta-\hat q_j)},
\end{equation}
where the $a_j^\beta$ are to be determined. Next, narrow down the choice of $a_j^\beta$ by setting
\begin{equation}
\label{0-3}
a_j^\beta= t \tilde a_j^\beta,
\end{equation}
where $\tilde a_j^\beta$ is either one or negative one. To satisfy (\ref{0-10}) we need $n_\beta$ to be even (one more
point can be easily inserted if the actual
$n_\beta$ is odd); in addition, the number of positive and negative $\tilde a_j^\beta$ must be the same.

%Further, by (\ref{c2}), and taking into account that $\hat q_j$ are symmetrically spaced in $I_\beta$, we see that (\ref{0-10})
%will hold if
%\begin{equation}
%\label{0-3}
%\sum_{j\in J_\beta} \delta v^\beta_j=0.
%\end{equation}

%We next show that $\kappa^2$ can be prescribed so that (\ref{0-3}) and energy conservation are satisfied.
To
simplify further calculations, we write conservation of energy (\ref{en-cons}) in the form
\begin{equation}
\label{0-11}
 \sum_{j\in J_\beta} (\delta v^\beta_j)^2=  K_\beta,
\end{equation}
where $K_\beta$ are any numbers satisfying
\begin{equation}
\label{kbeta}
K_\beta>0,\hspace*{0.5cm}\sum_{\beta=1}^B K_\beta=\frac{2 N}{M}\left(
{\mathcal E}-{\mathcal U}(\widehat{Q})-
\frac 12 \frac{M}{N} \sum_{\beta=1}^B (\overline{v}^\eta_\beta)^2 n_\beta
\right).
\end{equation}

Our goal now is to show that there is a choice of  $K_\beta$, $\hat \kappa^2$ and $t$ such that
$\delta v^\beta_j$ defined by (\ref{0-31}), (\ref{0-3}) satisfy equations
(\ref{0-4}), (\ref{0-11}), and (\ref{kbeta}).
Inserting (\ref{0-31}) into (\ref{0-4}) and (\ref{0-11}) yields,
respectively,
\begin{eqnarray}
t^2 \sum_{j\in J_\beta}\frac{1}{\psi_\eta(x_\beta-\hat q_j)} & = & \hat \kappa^2,\label{0-14}\\
t^2 \sum_{j\in J_\beta}\frac{1}{(\psi_\eta(x_\beta-\hat q_j))^2} & = & K_\beta.\label{0-15}
\end{eqnarray}
Combining these equations we get
\begin{equation}
\label{0-16}
t^2=\hat \kappa^2
\left(
\sum_{j\in J_\beta} \frac{1}{\psi_\eta(x_\beta-\hat q_j)}
\right)^{-1},
\end{equation}
%%%%%%%%%%%%%%%%%%%%%%%%%%%%%%%%%%%%%%%%
%%%%%%%%%%%%%%%%%%%%%%%%%%%%%%%%%%%%%%%
\begin{equation}
\label{0-17}
\hat \kappa^2
\sum_{j\in J_\beta} \frac{1}{(\psi_\eta(x_\beta-\hat q_j))^2}
\left(
\sum_{j\in J_\beta} \frac{1}{\psi_\eta(x_\beta-\hat q_j)}
\right)^{-1}=K_\beta.
\end{equation}
Substituting into (\ref{kbeta}) yields the choice of $\hat \kappa$:
\begin{equation}
\label{kapa}
\hat \kappa^2=
\frac{2 N}{M}\left(
{\mathcal E}-{\mathcal U}(\widehat{Q})-
\frac 12 \frac{M}{N} \sum_{\beta=1}^B (\overline{v}^\eta_\beta)^2 n_\beta
\right)
%%%%%%%%%%%%%%%%%%%%%%%%
\left(
\sum_{\beta=1}^B
\frac{
\sum_{j\in J_\beta}
\frac{1}{(\psi_\eta(x_\beta-\hat q_j))^2}
}
{
\sum_{j\in J_\beta} \frac{1}{\psi_\eta(x_\beta-\hat q_j)}
}
\right)^{-1}.
\end{equation}
The choice of all constants now should be made as follows:\newline
\noindent
1) Given ${\mathcal E}, \overline{v}^\eta_\beta, \hat q_j$, find $\hat \kappa$ by (\ref{kapa});\newline
\noindent
2) Determine $K_\beta$ from (\ref{0-17});\newline
\noindent
3)  Determine $t$ from (\ref{0-16});\newline
\noindent
4) Choose $\delta v_j^\beta$ by (\ref{0-31}), (\ref{0-3}).\newline
\noindent
Note that step 4 introduces non-uniqueness, but we are concerned only with existence of suitable velocity perturbations.
The actual choice of $\delta v_j^\beta$ will not change the mesoscopic discretization of momentum balance equation. Indeed,
once $\hat q_j, \hat v_j$ are chosen, we can approximate the integral in (\ref{0-8}) (for $x$ at the mesoscale mesh nodes)
by a Riemann sum corresponding
to the partition $\hat q_j$ of  $(0, L)$:
\begin{eqnarray}
\label{0-18}
T^\eta_{(c), 0}(t, x_\alpha)
& =  &
-\sum_{\beta=1}^B \sum_{j\in J_\beta} \frac{L_\eta}{n_\beta}
(
\delta v^\beta_j
)^2
\psi_\eta (x_\alpha-\hat q_j) \overline{\rho}^\beta\\
& = &
- \sum_{j\in J_\alpha}
(
\delta v^\alpha_j
)^2
\psi_\eta (x_\alpha-\hat q_j) \nonumber\\
& = &
-\hat \kappa^2,\hspace*{0.8cm}
\alpha=1, 2, \ldots, B. \nonumber
\end{eqnarray}
Therefore, the mesoscale mesh values of $T^\eta_{(c), 0}$ are all equal. This implies that a finite difference approximation
of $\partial_x  T^\eta_{(c), 0}$ on the mesoscale mesh vanishes.
The conclusion is that for isothermal dynamics, any suitable choice of a velocity perturbation produces a convective stress
that has zero divergence on the mesoscale.
%%%%%%%%%%%%%%%%%%%%%%%%%%%%%
\subsection{Zero-order isothermal continuum model}
Combining the approximation $\partial_x T^\eta_{(c), 0}=0$ with (\ref{0-8}), (\ref{0-81})
we obtain an isothermal zero-order continuum model
\begin{eqnarray}
\partial_t \overline{\rho}^\eta+\partial_x (\overline{\rho}^\eta\overline{v}^\eta) &=& 0,\label{c8-1}\\
\partial_t( \overline{\rho}^\eta\overline{v}^\eta)+\partial_x
\left(
\overline{\rho}^\eta(\overline{v}^\eta)^2
\right)
-\partial_x T^\eta_{(int), 0}& =& 0,\label{c9-1}
\end{eqnarray}
where $T^\eta_{(int), 0}$ is given by an integral expression (\ref{0-81}) (or by a discretization (\ref{0-71})). We can interpret
interaction stress as pressure. Then (\ref{0-81}) provides dependence of pressure on density,
which is non-local in space and non-linear. For small $h$ (large $N$), it can be approximated by
$$
T^\eta_{(int), 0}\approx-
\int_0^L U^\prime
\left(
\frac{M}{\overline{\rho}^\eta(t, y)}
\right)
\psi_\eta
\left(
x-y
\right)
dy,
$$
which is still non-local.
In the limiting case $\eta\to 0$, observing that convolution with $\psi_\eta$ is an approximate identity, this equation
can be reduced a local equation of state
$$
T^\eta_{(int), 0}\approx-
U^\prime
\left(
\frac{M}{\overline{\rho}^\eta(t, x)}
\right).
$$
If $\overline{\rho}^\eta$ is nearly constant, this equation can be linearized to produce a classical gas dynamics linear
equation of state. This shows that zero-order closure (\ref{0-81}) generalizes several classical phenomenological equations of
state. The connection between micro- and mesoscales is made explicit in (\ref{0-81}).
Using  higher order closure approximations, one can obtain other non-classical continuum models worth further
investigation.
%%%%%%%%%%%%%%%%%%%%%%%%%%%%%%%%%%%%%%%%%
%%%%%%%%%%%%%%%%%%%%%%%%%%%%%%%%%%%%%%%%%%%%%%%
%%%%%%%%%%%%%%%%%%%%%%%%%%%%%%%%%%%%%%%%%%%%%%%%%%%%
\section{Computational results}
\label{sect:computing}
%\subsection{Example: nonlinear acoustics of nearly rigid spheres}

In this section, the method developed in the previous sections is tested for a chain of $N=10,000$ to $N=80,000$ particles interacting with a
non-linear finite rage potential
\begin{equation}
\label{U}
U(\xi)=
\left\{
\begin{array}{ll}
C_r\left(\frac{1}{1-p}\xi^{1-p}x_\star-\xi x_\star^{1-p}+\frac{p}{p-1}x_\star^{2-p}\right), & \quad {\rm if}~\xi\in (0, x_\star]\\
0,  & \quad {\rm if}~\xi > x_\star\\
\end{array}
\right.
\end{equation}
where $p>1$, $x_\star=\alpha L$,  $\alpha\approx 1$ and $C_r$ is material stiffness. This potential mimics a Hertz potential used in modeling
of granular media. Particles in this model are centers of lightly touching spherical granules arranged in a chain, and the
ODEs model
acoustic wave propagation in this chain. The microscale equations are (\ref{ode-1}), (\ref{ode-2}) with initial conditions given below. The forces include
the pair interaction forces defined by $U$, and the exterior confining forces
acting on the first and last particles. The parameters of $U$ are chosen so that all particles stay within the interval $[0, L]$ for the duration of a simulation.
%
%\begin{figure}[h] \centerline{
%\includegraphics[width=6.0in,angle=0]{Figures_MCM/velocity_v_vbar_N40000_B50.pdf}
%} \caption{$N=40,000$, $B=50$. Dashed line: exact velocity $v_j$, $j=1,2,\ldots,N$; solid line: average velocity $\overline{v}_\beta$, $\beta=1,2,\ldots,B$.}
%\label{FB1}
%\end{figure}
%
To ensure that particles do not leave the interval  $[0, L]$, its endpoints are modeled as stationary particles that interact with the moving particles with forces generated by the same potential $U$.
If needed, stiffness of walls can be increased by using a different value of $C_r$.

%Let $x_\beta=\frac 12 \eta L+(\beta-1)\eta L$, $\beta=1, 2, \ldots, B$, be the centers of the mesocells $I_\beta$, where $\eta=\frac 1B$. The mesocell $I_\beta$ is defined as
%$
%I_\beta=[x_\beta-\frac 12 \eta L, x_\beta+\frac 12 \eta L).
%$
Let $x_\beta$ be the centers of the mesocells  $I_\beta$, $\beta=1, 2, \ldots, B$, as defined in Section \ref{sect:zero-closure}.
Next, let a window function $\psi(x)$ be the characteristic function of the interval $[-\frac 12 \eta L, \frac 12 \eta L)$.
The average density and momentum are defined by
(\ref{density}) and (\ref{mom}), respectively, and the average velocity is
$$
\overline{v}_\beta(t)=\frac{\sum_{j=1}^Nv_j(t)\psi(x_\beta-q_j(t))}{\sum_{j=1}^N\psi(x_\beta-q_j(t))}.
$$
\begin{figure}[h] \centerline{
\includegraphics[width=6.0in,angle=0]{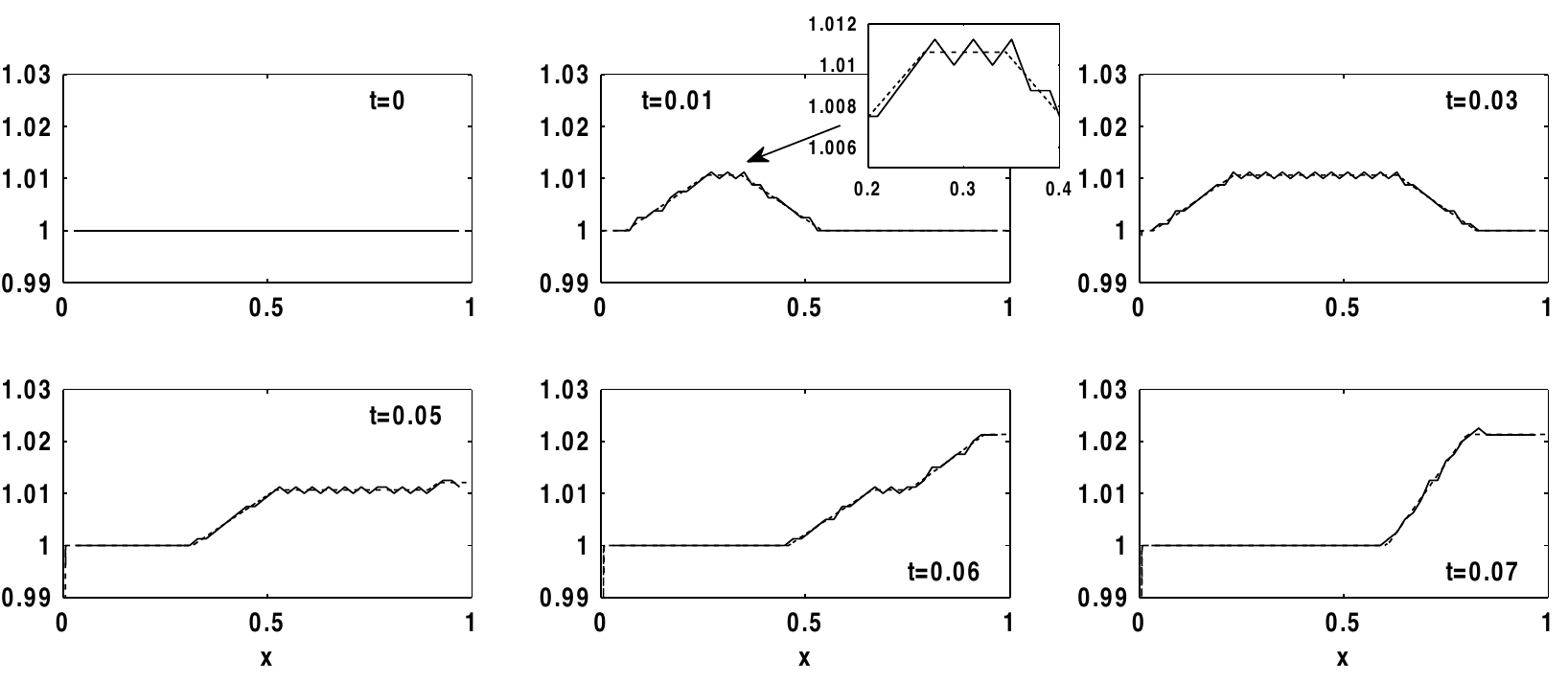}
} \caption{$N=40,000$, $B=50$. Dashed line: Jacobian $J(t,x_\beta)$; solid line: its mesoscale approximation $\frac LM \overline{\rho}^\eta(t,x_\beta)$, $\beta=1,2,\ldots, B$ according to  (\ref{0-1}). Blowup of results at $t=0.01$ shows discrepancy between  $J(t,x_\beta)$ and $\frac LM \overline{\rho}^\eta(t,x_\beta)$.}
\label{FB2}
\end{figure}

The average density $\overline{\rho}^\eta$ evaluated at the center $x_\beta$  of a mesocell $I_\beta$ can be written as
\begin{equation}\label{density_aver}
\overline{\rho}^\eta_\beta(t)=\overline{\rho}^\eta(t, x_\beta)=\frac{M}{N}\sum_{j=1}^N\frac{1}{\eta L}\psi\left(\frac{x_\beta-q_j(t)}{\eta}\right)=\frac{B}{N} \frac{M}{L} \sum_{j=1}^N\psi\left(\frac{x_\beta-q_j(t)}{\eta}\right)
\end{equation}
that shows that average density is proportional to the scale separation $B/N$.
\begin{figure}[h] \centerline{
\includegraphics[width=6.0in,angle=0]{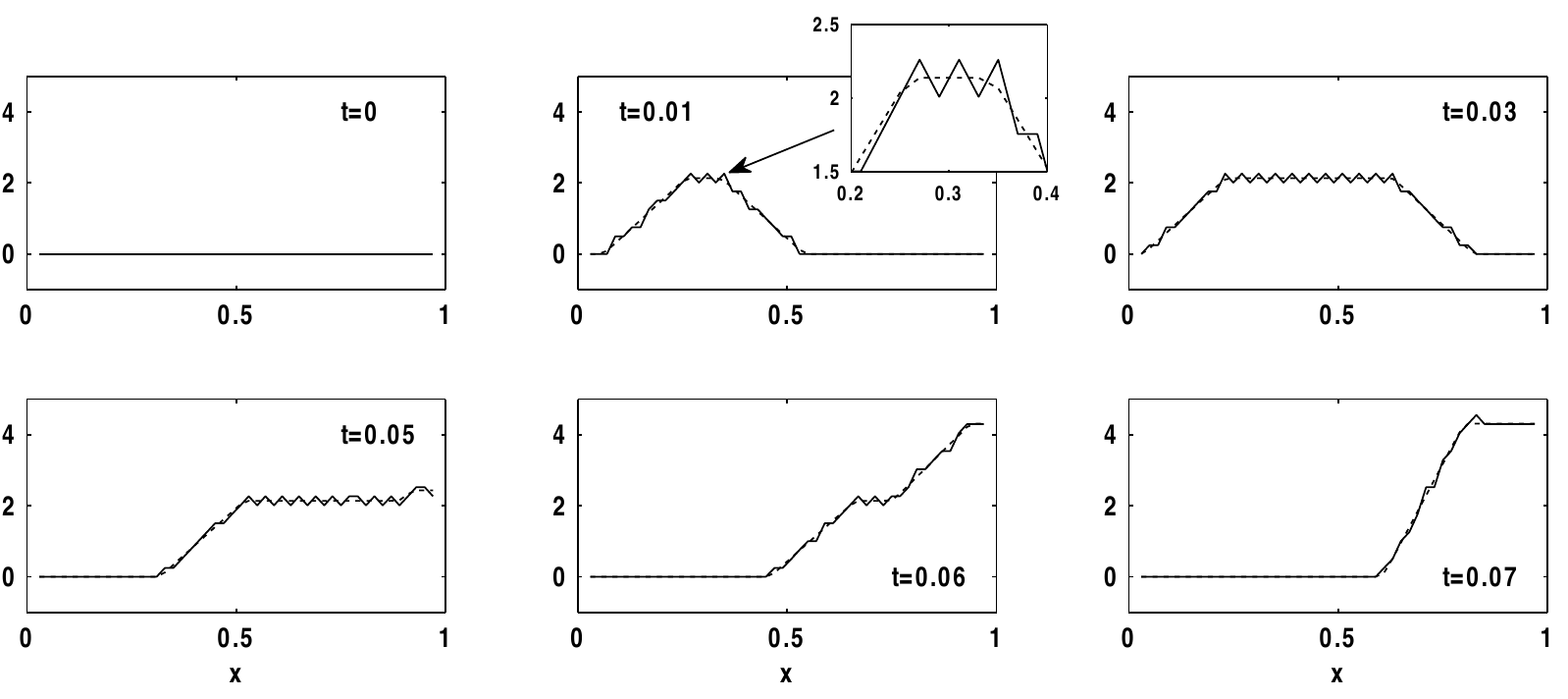}
} \caption{$N=40,000$, $B=50$. Dashed line: exact interaction stress $T^\eta_{(int)}(t,x_\beta)$; solid line: its approximation $T^\eta_{(int),0}(t,x_\beta)$, $\beta=1,2,\ldots, B$, defined in (\ref{0-81}). Blowup of results at $t=0.001$ shows difference between exact stress and its approximation.}
\label{FB4}
\end{figure}

We solve microscopic equations (\ref{ode-1}), (\ref{ode-2}) subject to the initial positions
\[
q_j^0=\left(j-\frac 12\right)h, \quad j=1, 2, \ldots, N, \quad h=\frac LN,
\]
and
%In our example, we let $q_j^0=(j-\frac 12)h$, $j=1, 2, \ldots, N$, where  microstep $h$ is defined in (\ref{h}). Note that $q_1(0)=\frac h2$,
%$q_N(0)=L-\frac h2$.
the initial velocities %are defined as
\[
v_j^0=
\left\{
\begin{array}{l}
\gamma, \hspace{59pt}  \mbox{ if } \ 0\leq q_j^0 \leq \frac L5, \\[5pt]
\gamma \left(-\frac 5L q_j^0+2 \right), \  \mbox{ if } \ \frac L5\leq q_j^0 \leq \frac{2L}{5}, \\[5pt]
0, \hspace{59pt}  \mbox{ if } \ \frac{2L}{5} \leq q_j^0 \leq L
\end{array}
\right.
\]
using the Velocity Stormer-Verlet method.
%The initial velocity is shown in Fig. \ref{FB1} in the top left panel.
We use $L=1$, $p=2$, $\alpha=1$, $\gamma=0.3$ and $C_r=100$. This velocity profile initiates an acoustic wave that propagates to the right. The stiffness constant $C_r$ can be used to ensure that particles have only small displacements from their equilibrium positions. Using initial velocity with higher $\gamma$ would require a higher value of $C_r$ to enforce smallness of typical particle displacements.

With fixed $N=40,000$, $B=50$, we integrate microscopic equations  (\ref{ode-1}), (\ref{ode-2}) until the acoustic wave reaches the right wall, interacts with it  and is about of being reflected to the left, which corresponds to $t=0.07$.
%With the time step $\Delta t=5\cdot 10^{-8}$ we integrate until $t_f=0.07$.
%Snapshots of quantities of interest are presented at times $t=0$, $0.01$, $0.03$, $0.05$, $0.06$ and $0.07$ to capture the most interesting dynamics.
To capture the most interesting dynamics, we present snapshots of results  at times $t=0$, $0.01$, $0.03$, $0.05$, $0.06$ and $0.07$.
To test our closure method, we compute microscopic positions $q_j$ and velocities $v_j$, $j=1,\ldots,N$, at every time step  and use them to evaluate primary mesoscopic variables:
%mesoscopic quantities at centers $x_\beta$ of mesocells:
average density $\overline{\rho}^\eta_\beta$  and average velocity $\overline{v}^\eta_\beta$, at mesocell centers $x_\beta$, $\beta=1,\ldots,B$.  These mesoscopic quantities are defined in  (\ref{density}), (\ref{mom}) (see also
(\ref{c6})).
%These mesoscopic quantities are defined in
%
%
%In Fig. \ref{FB1}, we compare microscopic velocities $v_j$, $j=1,2,\ldots,N$ with the average velocities $\overline{v}^\eta_\beta$, $\beta=1,2,\ldots,B$ and $B=50$. The initial micro- and mesoscopic velocity profiles are shown in the top left panel of Fig. \ref{FB1}. "Exact" microscopic quantites are plotted using a dashed line while mesoscopic quantities are plotted with a solid line. We see that micro- and mesoscale velocities agree very well.
%%
%
 They are then employed
 in computing of the zero-order approximation $T^\eta_{(int),0}(t,x_\beta)$ defined in (\ref{0-81}). We compare this mesoscopic approximation %at every time step
 with the ``exact" microscopic
 %analogue,
 interaction stress $T^\eta_{(int)}(t,x_\beta)$ defined in  (\ref{c2}), and also test other approximations given by (\ref{0-1}), (\ref{0-2}).

%In Fig. \ref{FB1}, we show microscopic velocities $v_j$, $j=1,\ldots,N$, together with  mesoscopic velocities $\overline{v}^\eta_\beta$, $\beta=1,2,\ldots,B$.

Comparing $v_j$, $j=1,\ldots,N$ and  $\overline{v}^\eta_\beta$, $\beta=1,2,\ldots,B$ (not shown here) we find that micro- and mesoscale velocities are essentially indistinguishable during the simulation time.

% We see that micro- and mesoscale velocities are essentially indistinguishable.
%
\begin{figure}[h] \centerline{
\includegraphics[width=6.0in,angle=0]{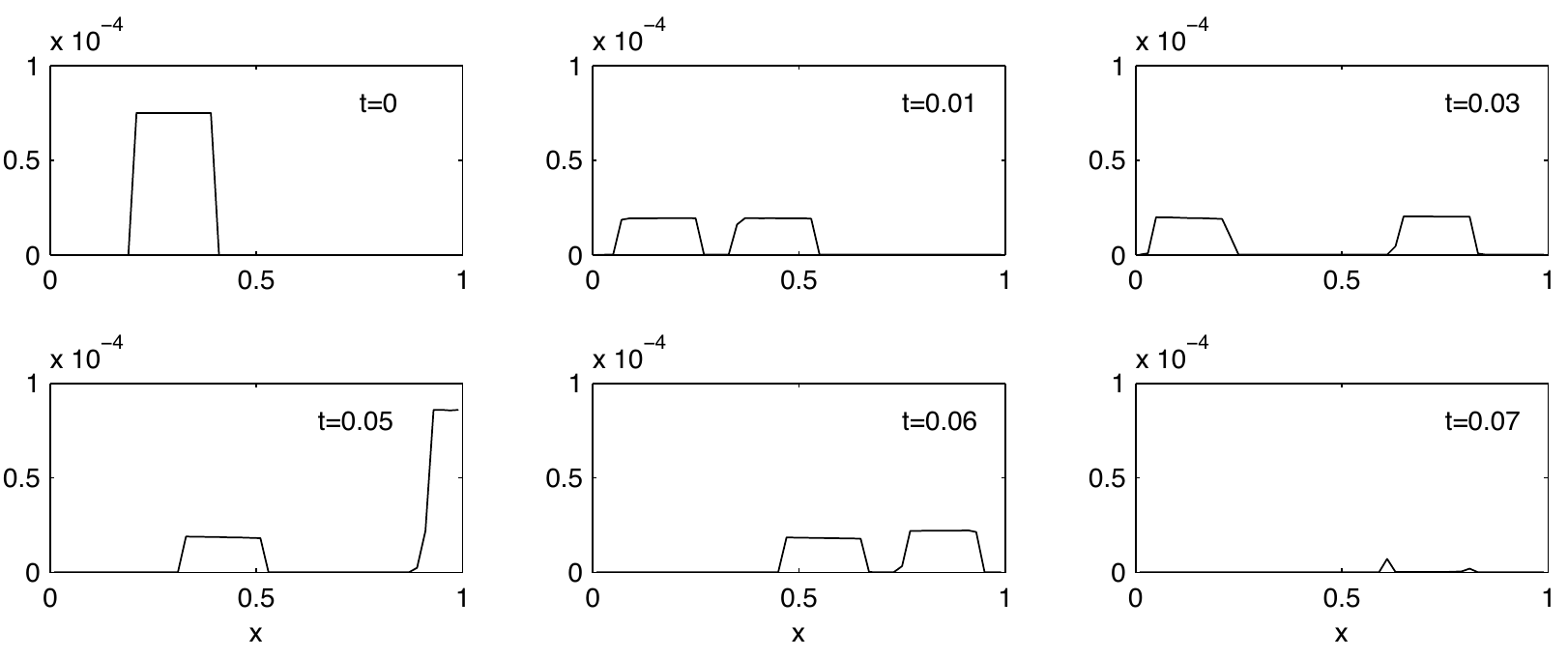}
} \caption{$N=40,000$, $B=50$. Convective stress ${T}^\eta_{(c)}(t,x_\beta)$, $\beta=1,2,\ldots, B$, defined in (\ref{0-8}).}
\label{FB5}
\end{figure}
In  Fig. \ref{FB2},  we analyze microscopic Jacobian  $J(t,x_\beta)$ together with  its zero-order mesoscopic approximation $\frac LM \overline{\rho}^\eta (t, x_\beta)$ obtained according to  (\ref{0-1}). In this and other figures, we plot
``exact" microscopic quantities using a dashed line while mesoscopic quantities are depicted with a solid line. Results shown in Fig. \ref{FB2} indicate that
$\frac LM\overline{\rho}^\eta(t,x_\beta)$  exhibits some oscillations whose amplitude is about $10^{-3}$ as compared to Jacobian $J(t,x_\beta)$.
The oscillations are likely caused by the choice of  a window function $\psi$.
For computational testing, we chose  $\psi$ to be a characteristic function. The main reason was to try ``the worst case scenario" concerning smoothness of $\psi$. We expected that this window function would produce more oscillations than a smoother $\psi$. A good agreement between our approximation and the direct simulation results strongly suggest that the proposed method is viable. We believe that it should perform better with a smoother choice of $\psi$.
The oscillations present in $\frac LM \overline{\rho}^\eta(t,x_\beta)$ are amplified
in the approximated stress ${T}^\eta_{(int),0}(t,x_\beta)$, due to  the rather high stiffness constant $C_r=100$, as shown in Fig. \ref{FB4}.
We also compare microscopic $J(t,x_\beta)\tilde v(t,x_\beta)$ with its zero-order approximation $\frac LM \overline{\rho}^\eta (t, x_\beta)\overline{v}^\eta (t, x_\beta)$ according to (\ref{0-2}). Graphs are not shown here but we find that these quantities agree very well similar to micro- and mesoscale velocities. %plotted in Fig.  \ref{FB2}.
This is expected since the average density is approximately identity with small oscillations.
Finally, we verify that the dynamics is quasi-isothermal by plotting the convective stress $T^\eta_{(c)}(t,x_\beta)$ defined in (\ref{c11})  in Fig. \ref{FB5}. As can be seen, fluctuations in the convective stress do not exceed $10^{-4}$ throughout computational time, therefore, the kinetic energy of velocity fluctuations is small.
%Smallness of velocity fluctuations was also verified by comparing microscopic velocity interpolant with the average velocity.
%%

%We also compare "exact" microscopic quantities $J(t,x_\beta)$, $J(t,x_\beta)v(t,x_\beta)$ computed at the centers of mesocells $x_\beta$, $\beta=1,2,\ldots,B$, with their zero-order mesoscopic approximations $\frac LM \overline{\rho}^\eta (t, x_\beta)$, $\frac LM \overline{\rho}^\eta (t, x_\beta)\overline{v}^\eta (t, x_\beta)$ according to (\ref{0-1}), (\ref{0-2}), respectively, as well analyze the magnitude of velocity fluctuations using a convective stress $T^\eta_{(c)}(t,x_\beta)$ defined in (\ref{c11}).  In the graphs presented below, we compare cases with $B=50$ and $B=100$ to test the effect of scale separation. With $N=40,000$,
%

%%%%%%%%%%%%%%%%%%%%%%%%%%%%%%%%%%%
%
%%
%\begin{figure}[h] \centerline{
%\includegraphics[width=6.0in,angle=0]{Figures_MCM/Jac_density_Nvaries_i4.pdf}
%} \caption{$B=50$, $N$ varies, $t=0.004$. Dashed line: Jacobian $J(t,x_\beta)$; solid line: its mesoscale approximation $\frac LM \overline{\rho}^\eta(t,x_\beta)$, $\beta=1,2,\ldots, B$}
%\label{FBJDNvaries}
%\end{figure}
%%
%
%%
%\begin{figure}[h] \centerline{
%\includegraphics[width=6.0in,angle=0]{Figures_MCM/T2_T2bar_Nvaries_i4.pdf}
%} \caption{$B=50$, $N$ varies, $t=0.004$. Dashed line: exact interaction stress $T^\eta_{(int)}(t,x_\beta)$; solid line: its mesoscale approximation $T^\eta_{(int),0}(t,x_\beta)$}
%\label{FBT2Nvaries}
%\end{figure}
%%

%%%%%%%%%%%%%%%%%%%%%%%%%%%%%%%%%%

%
\begin{figure}[h] \centerline{
\includegraphics[width=6.0in,angle=0]{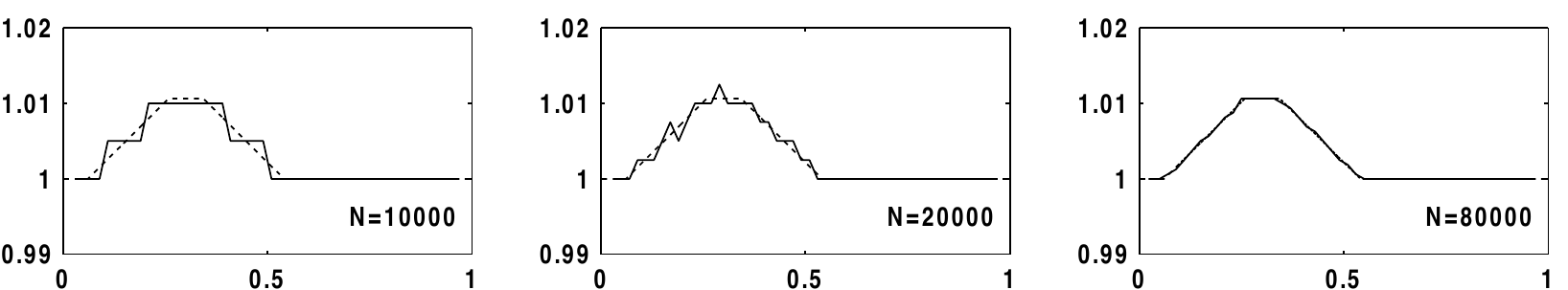}
} \caption{Effect of the scale separation on $\frac LM \overline{\rho}^\eta$. $B=50$ is fixed, $N$ varies, data is taken at the same $t=0.01$. Dashed line: Jacobian $J(t,x_\beta)$; solid line: its mesoscale approximation $\frac LM \overline{\rho}^\eta(t,x_\beta)$, $\beta=1,2,\ldots, B$.}
\label{FBJDNvaries}
\end{figure}

\begin{figure}[h] \centerline{
\includegraphics[width=6.0in,angle=0]{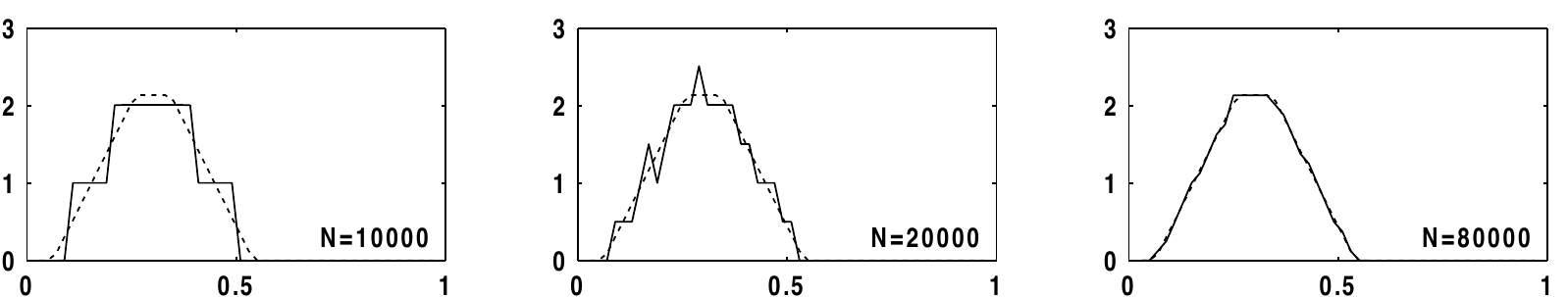}
} \caption{Effect of the scale separation on $T^\eta_{(int),0}$. $B=50$ is fixed, $N$ varies, data is taken at the same $t=0.01$. Dashed line: exact interaction stress $T^\eta_{(int)}(t,x_\beta)$; solid line: its mesoscale approximation $T^\eta_{(int),0}(t,x_\beta)$, $\beta=1,2,\ldots, B$.}
\label{FBT2Nvaries}
\end{figure}

We next tested the effect of the scale separation on the quality of the zero-order approximation. With fixed $B=50$, we allowed $N$ vary from $10,000$ to $80,000$ and followed the evolution of mesoscale quantities of interest: $\frac LM \overline{\rho}^\eta(t,x_\beta)$ and $T^\eta_{(int),0}(t,x_\beta)$. Snapshots of these functions at the same representative time $t=0.01$ are plotted in Figs. \ref{FBJDNvaries} and \ref{FBT2Nvaries}, respectively, with $N=10,000$, $N=20,000$ and $N=80,000$.
%All other parameters are kept the same.
The results with $N=40,000$ at the same time are given the middle top panels in Figs. \ref{FB2}, \ref{FB4} for comparison. It is clear that as scale separation increases, oscillations in both $\frac LM \overline{\rho}^\eta(t,x_\beta)$ and $T^\eta_{(int),0}(t,x_\beta)$ diminish and when $N=80,000$, the exact microscopic quantities and their mesoscale approximations are almost indistinguishable.
%

%%%%%%%%%%%%%%%%%%%%%%%%%%%%%%%%%%

%
\begin{figure}[h] \centerline{
\includegraphics[width=2.9in,angle=0]{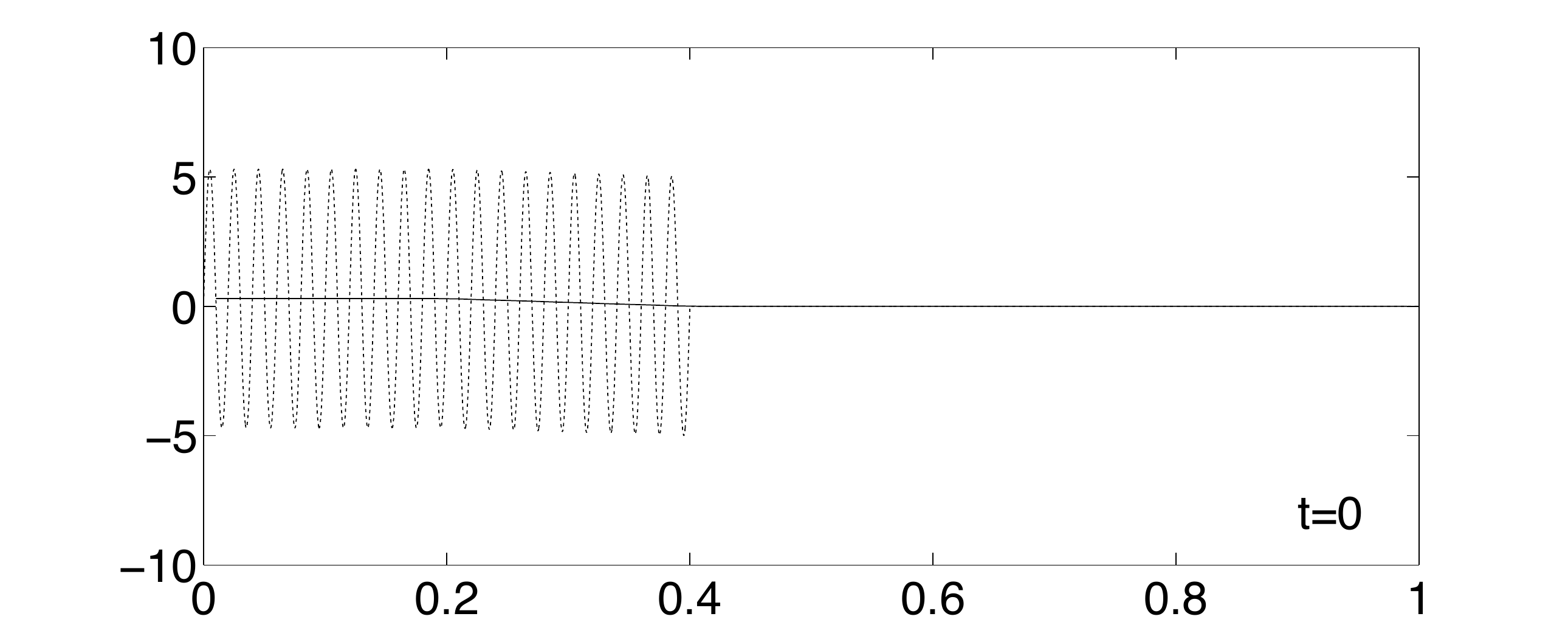}
\includegraphics[width=3.6in,angle=0]{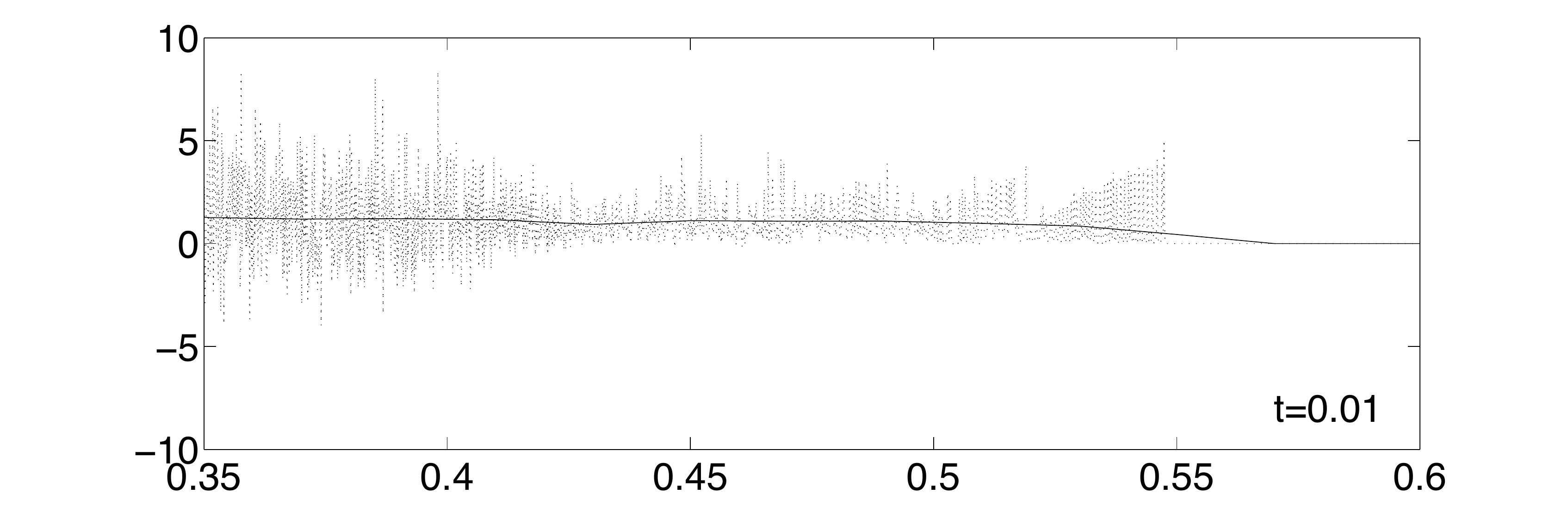}
} \caption{Example with imposed high frequency oscillations. $N=10,000$, $B=50$. Dashed line: exact velocity $v_j$, $j=1,2,\ldots,N$; solid line: average velocity $\overline{v}_\beta$, $\beta=1,2,\ldots,B$.}
\label{FB11}
\end{figure}
\begin{figure}[h] \centerline{
\includegraphics[width=2.9in,angle=0]{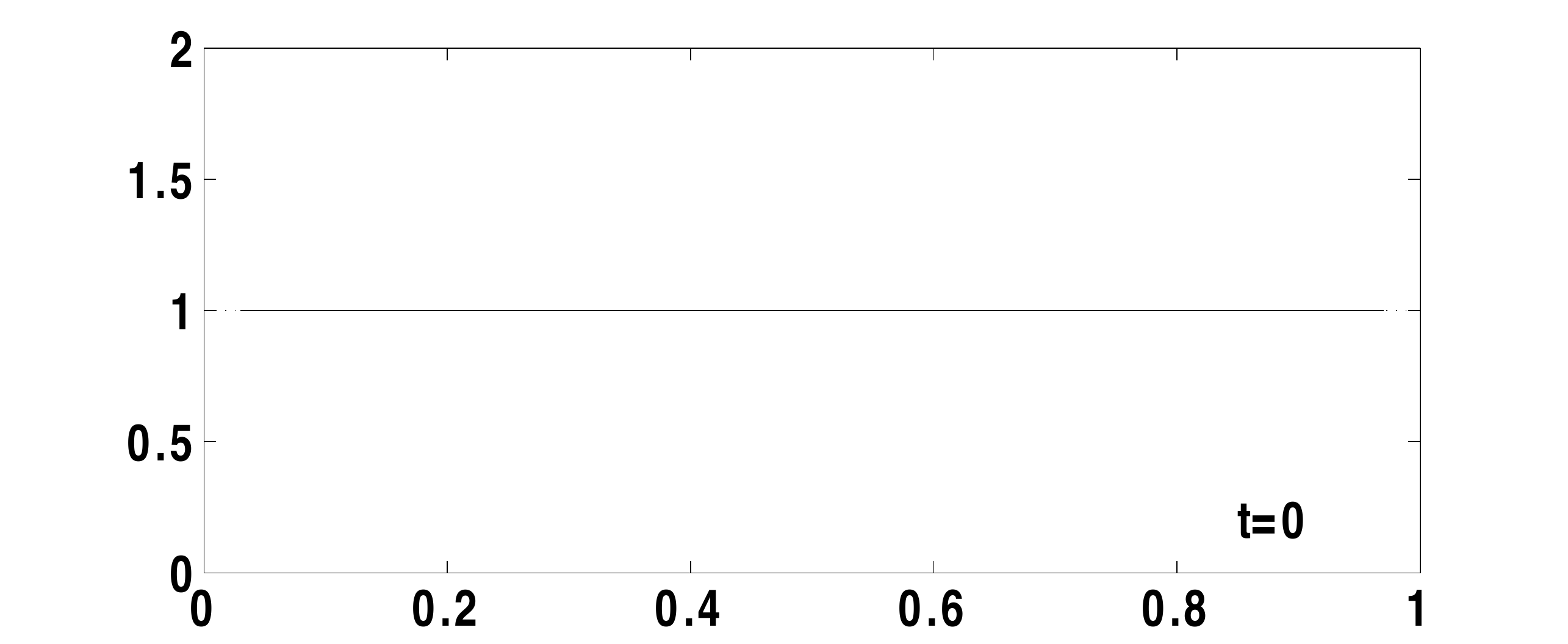}
\includegraphics[width=3.6in,angle=0]{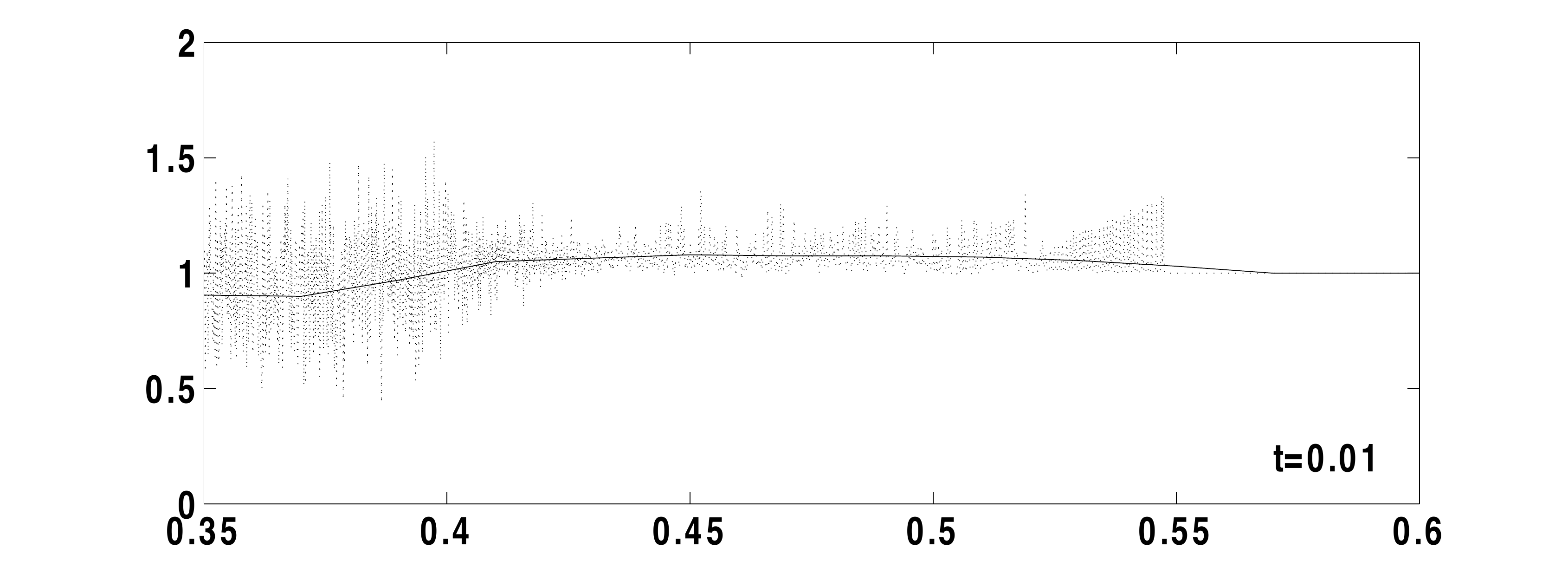}
} \caption{Example with imposed high frequency oscillations. $N=10,000$, $B=50$. Dashed line: Jacobian $J(t,x_\beta)$; solid line: its mesoscale approximation $\frac LM \overline{\rho}^\eta(t,x_\beta)$, $\beta=1,2,\ldots, B$ (compare with Fig. \ref{FB2}).}
\label{FB13}
\end{figure}

In the above example, fluctuations of microscopic velocities about their average values were very small and the zero-order approximation worked well.
Next we show that if microscopic velocities have high fluctuations then the zero-order approximation is not capable of captioning an appropriate dynamics.

We demonstrate this by imposing high frequency $k$ oscillations with relatively large amplitude $a$ on the nonzero portion of the initial velocity used in the previous experiments. The initial velocity is
\[
v_j^0=
\left\{
\begin{array}{l}
\gamma + a\sin(\frac{5k\pi}{L} q_j^0), \hspace{57pt}  \mbox{ if } \ 0\leq q_j^0 \leq \frac L5, \\[5pt]
\gamma \left(-\frac 5L q_j^0+2\right)+a\sin(\frac{5k\pi}{L} q_j^0) , \  \mbox{ if } \ \frac L5\leq q_j^0 \leq \frac{2L}{5}, \\[5pt]
0, \hspace{122pt}  \mbox{ if } \ \frac{2L}{5} \leq q_j^0 \leq L.
\end{array}
\right.
\]
and it is plotted in the left panel of Fig. \ref{FB11}. We use $a=5$ and $k=20$ that gives one period of imposed oscillations per mesocell. This microscopic initial velocity has a property that the average velocity at time $t=0$ is exactly the same as in the previous example.
%s used, so that we can investigate the effect of velocity fluctuations.
Simulations were done with $N=10,000$ until the same $t=0.07$.

\begin{figure}[h] \centerline{
\includegraphics[width=6.0in,angle=0]{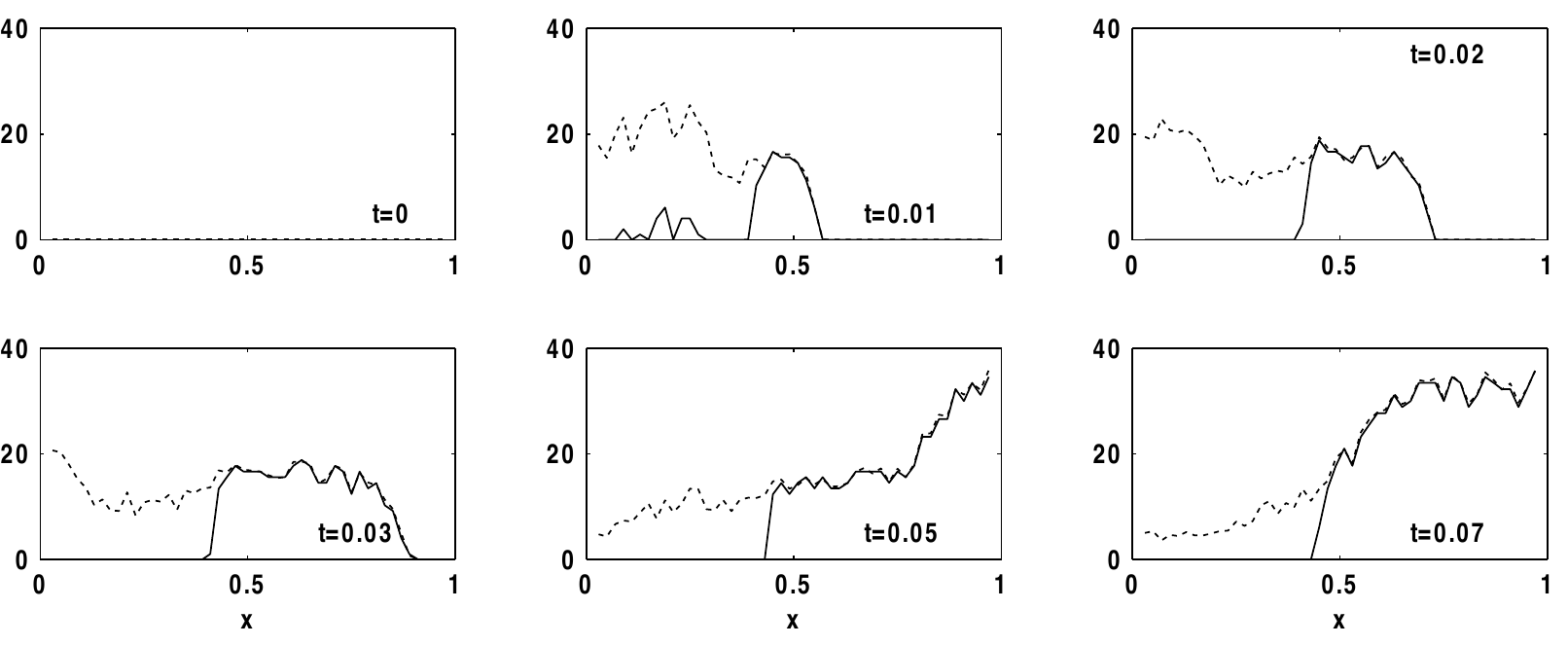}
} \caption{Example with imposed high frequency oscillations. $N=10,000$, $B=50$. Dashed line: exact interaction stress $T^\eta_{(int)}(t,x_\beta)$; solid line: its mesoscale approximation $T^\eta_{(int),0}(t,x_\beta)$ (compare with Fig. \ref{FB4}).}
\label{FB14}
\end{figure}

The right panel of
Fig. \ref{FB11} shows a typical microscopic velocity profile together with its average velocity (taken at $t=0.01$):  to the left from the wave front, the microscopic velocity has large frequency oscillations (due to dispersion?) with an amplitude sometimes exceeding the initial amplitude by a factor of $1.5$  and to the right from the wave front, the microscopic velocity is zero. Clearly, the average velocity is very different from the microscopic velocity.
%
% propagation of imposed fluctuations in velocity to the right. We also note the effect of dispersion here since the amplitude of oscillations decreases with time but oscillations are present eventually on the entire interval $[0,L]$.
%(SHOULD WE WRITE ABOUT DISPERSION HERE?).

Analysis of microscopic Jacobian $J(t,x_\beta)$ and mesoscopic $\frac LM \overline{\rho}^\eta(t,x_\beta)$ reveals that these functions have qualitatively the same dynamics as micro- and mesoscale velocities, respectively, shown in Fig. \ref{FB11}. We plot the former in Fig. \ref{FB13} where the left panel has graphs at $t=0$ while the right panel shows typical structure with data taken at $t=0.01$.  It is interesting to note that the zero-order approximation  $T^\eta_{(int),0}(t,x_\beta)$ to the  interaction stress  $T^\eta_{(int)}(t,x_\beta)$
plotted in Fig. \ref{FB14} does not agree in those areas that were affected by large magnitude oscillations in microscopic velocities while agrees well in those areas to which oscillations have not come yet. This finding suggests that indeed the zero-order approximation should not be used for large frequency oscillations in microscopic velocities and a higher order approximation is needed.
%It should be noted that in the areas that we not affected by large amplitude fluctuations, the zero-order approximation actually works really well and oscillations mentioned in the previous examples are not present. This is perhaps due to the fact that the variation in stress is much bigger than in the case considered previously and averaged stress is more sensitive to large changes.

%
\begin{figure}[h] \centerline{
\includegraphics[width=6.0in,angle=0]{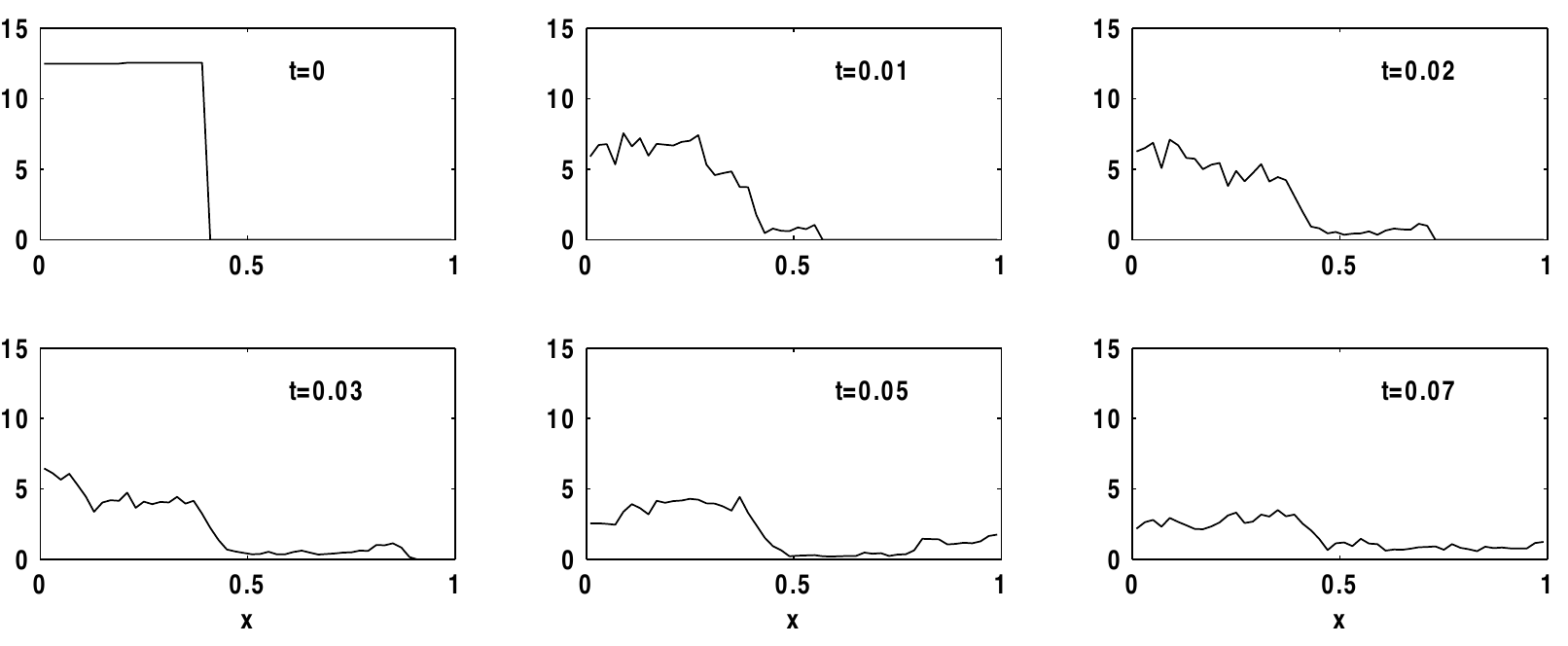}
} \caption{Example with imposed high frequency oscillations. $N=10,000$, $B=50$. Convective stress ${T}^\eta_{(c)}(t,x_\beta)$, $\beta=1,2,\ldots, B$ (compare with Fig. \ref{FB5}).} %!!
\label{FB15}
\end{figure}

Finally, in Fig. \ref{FB15} we plot the convective stress ${T}^\eta_{(c)}(t,x_\beta)$ whose large values confirm that oscillations in microscopic velocities are much bigger during the computational time than those in the first example.
When the initial velocity has  fluctuations with frequency higher than $k=20$, discrepancy between micro- and mesoscale quantities is even more pronounced.

%%%%%%%%%%%%%%%%%%%%%%%%%%%%%%%%%%%%%%%%%%%%%

%%%%%%%%%%%%%%%%%%%%%%%%%%%%%%%%%%%%%%%%%%%%%%%%%%%%%
\section{Zero-order closure: applicability}
\label{sect:applicability}
Zero-order closure is very similar to the use of the Cauchy-Born rule in quasi-continuum simulations
of solids. Here, the nodes of the
mesoscale mesh can be thought of as  "representative particles". These particles are moved with the average
velocity, while the velocities of other particles are assigned by interpolation. A construction of an interpolant should
take into account the physics of the microscopic model such as energy conservation. In the computational example of
Section \ref{sect:computing}, zero-order approximation turns out to be quite accurate, when non-oscillatory initial conditions are imposed. In this case, we found that
approximate and exact stresses agree rather well, and this agreement becomes better with increasing scale separation.

%This is why we developed the interpolant that produces an energy preserving collection of approximate particle
%positions and velocities, compatible with given average density and momentum.

This does not mean that zero-order closure always works well. Our numerical simulations suggest that applicability of
zero-order closure is determined by initial conditions, exterior forces and
interaction potential (arranged in order or importance).

Approximating functions by their averages we neglect fluctuations. Therefore,
the initial velocities should have small fluctuations. Initial positions should be chosen so that the number
of particles in a meso-cell varies slightly from one cell to another. The initial velocity fluctuations in our first example are small, and
convective stress at later times is by three orders of
magnitude
smaller than interaction stress. This remains true on the time interval sufficient  for the traveling wave to reach the opposite
end of the chain.

For one-dimensional problems, convective stress is proportional to the kinetic energy of
velocity
fluctuations. This kinetic energy  can be naturally associated with upscaling temperature. Relative smallness of the convective stress mens that upscaling temperature is nearly zero. Therefore,
the corresponding dynamics can be termed {\it cold}. We also note that cold dynamics is a special case of isothermal
dynamics, considered in Section \ref{sect:zero-closure}. As has been remarked earlier, isothermal
dynamics implies that divergence of
the convective stress is nearly zero on mesoscale, and thus can be neglected compared with the divergence of the
interaction stress.

Another consideration is related to inhomogeneity in actual particle distribution. In our example, deviations of about 4\%
in relative particle positions produced visible oscillations in the approximation of the interaction stress.  This amplification of small perturbations is due to the stiffness of the interaction potential. However, the same stiffness prevents particle aggregation, keeping the interparticle  distances bounded from below. Bounds from above are difficult to enforce with the chosen potential because it does not have a potential well. The isolated particle system with this potential would just fall apart. This phenomenon is common place for granular materials. The particles remain confined to the domain (container) only because they are repelled by the walls. Walls have very little direct influence on the interparticle distances in the systems' interior.
Therefore, applicability of zero-order closure also depends on the stiffness of the problem, and more generally on how well the potential enforces uniform particle distribution. In that sense, zero-order closure makes a reasonable approximation for lattice systems modeling small deformation of solids at constant temperature.

To further understand limitations of zero-order closure, consider the effect of increasing the order $n$ of the
Landweber approximations (\ref{seq1}).
The Fourier transform the kernel of $I-R_\eta$ is equal to
$$
1-e^{- \eta^2 \pi^2 \bxi \cdot \bxi }.
$$
It is very small for $\xi$ close to zero, and then increases to one as $|\xi|$ goes to infinity.
Therefore, $I-R_\eta$ acts as a
filter
damping low frequencies and thus emphasizing higher frequency content of the signal. Higher order  approximations amount to applying
convolutions
$\sum_{k=1}^n (I-R_\eta)^k$ to mesoscale averages. As $n$ increases, high frequency
content of the reconstruction will be increasingly amplified.
This suggests that systems capable of producing large fluctuations should be handled with higher order approximations.

A related comment is that averages of fluctuations can become additional state variables in a mesoscale continuum model. A familiar
example is the use of the averaged energy balance equation  (see \cite{mb} for derivation), in addition to the mass
and momentum balance.  The energy balance equation describes evolution of the density of kinetic energy of velocity fluctuations. An intriguing question here is how the model with just two equations of
balance but high order closure approximation compares with  a zero-order closure model containing all three
balance equations. In classical physics, additional balance equations are often introduced as a means of compensating for
errors introduced by replacing state variables with their averages. Use of higher order closure could offer an alternative
to this approach. Indeed, suppose that one is interested only in tracking density and velocity on mesoscale. The corresponding
two balance equations contain only two microscale quantities: velocity field $\tilde{v}$ and the Jacobian $J$ of the inverse
position map $\tilde{q}^{-1}$. If $\tilde{v}$ and $J$ can be accurately reconstructed from their averages,
we do not need to deal with the energy balance equation. This observation offers a new way of reducing computational cost.
Higher order approximation are more expensive than zero-order, but using more balance equations also increases
computational cost. We also note that increasing the order of closure approximations involves repeated convolutions with
the window function $\psi_\eta$.  On the other hand, simulating an energy balance involves numerical integration of an
additional non-linear integral-differential equation, a  much more difficult task.

%%%%%%%%%%%%%%%%%%%%%%%%%%%%%%%%%%%%%%%%%%%%%%%%%%%%%
\section{Conclusions}

We propose a closure method that gives closed form approximations for mesoscale continuum mechanical fluxes (such as stress) in terms of primary mesoscopic variables (such as average density and velocity). Our closure construction is
based on iterative regularization methods for solving first kind integral equations. Such integral equations are relevant because mesoscopic density and velocity are related to the corresponding microscopic quantities via a linear convolution operator. The problem of inverting convolution operators is unstable (ill-posed) and requires regularization. Use of the well known Landweber iterative regularization yields  successive approximations, of orders zero, one, two and so forth, to  interpolants of particle positions and velocities in terms of available averages. Closure is achieved by inserting any of these approximations into the equations for fluxes instead of the actual particle positions and velocities. Low order approximations are simpler to implement, while higher order approximations can be used to more accurately reproduce the high frequency content of the microscopic quantities.

The above general strategy is applied in the paper to spatially averaged dynamics of classical particle chains.
We focus on the simplest zero-order approximation and show numerically that it works reasonably well as long as  initial conditions have small velocity fluctuations.  The case of large fluctuations in velocities should be handled by higher order approximations.

%%%%%%%%%%%%%%%%%%%%%%%%%%%%%%%%%%%%%%%%%%%%%%%a%%
\section{Acknowledgments}
Work of
Alexander Panchenko was supported in part by DOE grant DE-FG02-05ER25709 and by NSF grant
OISE-0438765.
Work of R. P. Gilbert was supported in part by NSF grants OISE-0438765 and DMS-0920850, and by the Alexander v. Humboldt Senior Scientist Award at the Ruhr Universitat Bochum.

%%%%%%%%%%%%%%%%%%%%%%%%%%%%%%%%%%%%%%%%%%%%%%%%%%%%%%%%%%%%%

%\bibliography{references_meso}

%%%%%%%%%%%%%%%%%%%%%%%%%%%%%%%%%%%%%%%%%%%%%%%%%%%%%%%%%%%%%%%%
%%%%%%%%% E N D  OF   T H E  M A I N  T E X T %%%%%%%%%%%%%%%%%%%%%%%%%%%%%%%%%%%%%
%%%%%%%%%%%%%%%%%%%%%%%%%%%%%%%%%%%%%%%%%%%%%%%%%%%%%%%%%%%%%%%

\end{document}